\documentclass[aps,pre,twocolumn,floatfix,letterpaper,amssymb]{revtex4-2}
\usepackage{verbatim}
\usepackage{amsmath}
\usepackage{graphicx}
\usepackage{hyperref}
\usepackage{multirow}
\usepackage{xcolor}
\newcommand{\red}[1]{\textcolor{black}{#1}}
\newcommand{\newred}[1]{\textcolor{black}{#1}}
\usepackage{soul}

\begin{document}

\title{Disorder in dissipation-induced topological states: Evidence for
a different type of localization transition}

\author{Alon Beck and Moshe Goldstein}

\affiliation{Raymond and Beverly Sackler School of Physics and Astronomy, Tel
Aviv University, Tel Aviv 6997801, Israel}

\date{\today}
\begin{abstract}
The quest for nonequilibrium quantum phase transitions is often hampered by the tendency of driving and dissipation to give rise to an effective temperature, resulting in classical behavior. \red{Could this be different when the dissipation is engineered to drive the system into a nontrivial quantum coherent steady state?} In this work we shed light on this issue by studying the effect of disorder on recently-introduced dissipation-induced Chern topological states, and examining the eigenmodes of the Hermitian steady state density matrix or entanglement Hamiltonian. We find that, similarly to equilibrium, each Landau band has a single delocalized level near its center. However, using three different finite size scaling methods we show that the critical exponent $\nu$ describing the divergence of the localization length upon approaching the delocalized state is significantly different from equilibrium if disorder is introduced into the non-dissipative part of the dynamics. This indicates a different type of nonequilibrium quantum critical universality class accessible in cold-atom experiments.
\end{abstract}
\maketitle

\textit{Introduction.\textemdash}
Recent years have seen a surge of interest in the driven-dissipative dynamics of quantum many body systems~\cite{Kamenev2009,Sieberer2016}.
Of particular interest is the possibility of new nonequilibrium quantum critical phenomena.
However, typically far-from-equilibrium conditions give rise to an effective temperature governing the long time physics, and leading to classical criticality.
This stands in line with the usual perception of driving and dissipation as causing decoherence and destroying subtle quantum phenomena.
This point of view has been challenged by recent works showing how coupling to an environment could be engineered to drive a system towards desired steady states displaying quantum correlations~\cite{Diehl2008,Kraus2008,Verstraete2009,Weimer2010,Otterbach2014,Lang2015,Zhou2017}, such as nonequilibrium topological states~\cite{Diehl2011,Bardyn2012,Bardyn2013,Koenig2014,Kapit2014,Budich2015,Iemini2016,Gong2017,Goldstein2019,Shavit2020,Tonielli2020,Yoshida2020,Bandyopadhyay2020,Altland2020}.
In particular, Refs.~\cite{Goldstein2019,Shavit2020} introduced a protocol, realizable with cold atoms, for purely dissipative dynamics which approaches at a finite rate a mixed steady state as close as desired to a pure topological state.
Yet, the resulting topology is encoded in the \textit{Hermitian} steady state density matrix, 
giving rise to \red{the} same topological classes as in equilibrium~\cite{Diehl2011,Bardyn2012,Bardyn2013,Rivas2013,Huang2014,Viyuela2014,Nieuwenburg2014,Budich2015,Budich2015b,Grusdt2017,Bardyn2017,Bardyn2018,Zhang2018,Goldstein2019,Coser2019,Lieu2020,Yoshida2020,Altland2020}. Could this
\red{new type of engineered driving} still lead to new quantum nonequilibrium criticality?

Every natural system exhibits imperfections and disorder.
In equilibrium, it has long been recognized that disorder is actually essential for stabilizing the most basic topological phase, the integer quantum Hall state~\cite{Klitzing1980}.
Disorder localizes all states in a Landau level except one at energy $E_c$. The \red{wavefunction} localization length diverges as one approaches it as~\cite{Huckestein1995,Evers2008}
\begin{equation}
\xi(E) \sim \left|E-E_{c}\right|^{-\nu},\label{eq:localization_length}
\end{equation}
with a critical exponent $\nu$ governing the plateau transition.
Lately, a debate arose regarding the theoretical value of $\nu$~\cite{Slevin2009,Obuse2010,Amado2011,Fulga2011,Slevin2012,Obuse2012,Nuding2015,Gruzberg2017,Ippoliti2018,Puschmann2019,Zhu2019,Sbierski2021}, and its relation to experiment~\cite{Li2005,Li2009,Giesbers2009}; 
the currently accepted value 
is 2.5--2.6.

In this work we study the interplay between disorder and the recipe of Refs.~\cite{Goldstein2019,Shavit2020} for dissipatively-inducing Chern-insulator states, through the effects of disorder on the eigenmodes of the steady-state density matrix, which is experimentally measurable in cold atoms~\cite{Hauke2014,Flaschner2016,Tarnowski2017,Ardilla2018,Zheng2020}. This is thus a \emph{Hermitian} localization problem, unrelated to disordered nonhermitian Hamiltonians~\cite{Hatano1996,Ashida2020}.
We show that disorder in the system-bath coupling leads to the same universality class as in equilibrium, while disorder perturbing the system Hamiltonian is not. We employ three different finite size scaling (FSS) methods, based on
(a) the number of conducting states~\cite{Yang1996,Zhu2019}; (b) the local Chern marker~\cite{Bianco2011};
(c) the transfer matrix Lyapunov exponent~\cite{Evers2008,Puschmann2019}. The final results are presented
in Table~\ref{tab:nu_results}; all methods show that the out-of-equilibrium $\nu$ is larger by 0.5--0.6 than equilibrium, hinting at a different universality class.

\begin{table*}
	\begin{ruledtabular}
		\begin{tabular}{c|ccccccc||cccccccccc}
			& \multicolumn{7}{c||}{Equilibrium} & \multicolumn{10}{c}{Out of equilibrium}\tabularnewline
			Method & $W$ & Geometry & $L$ & $L_{x}$ & $N_{g}$ & $M$ & $L_{x}^{\mathrm{eff}}$ & $W$ & $\mu^{\mathrm{eff}}$ & $\frac{\gamma^{\mathrm{in}}}{\gamma^{0}}$ & Geometry & $L$ & $L_{x}$ & $p$ & $N_{g}$ & $M$ & $L_{x}^{\mathrm{eff}}$\tabularnewline
			\cline{2-18} 
			I & 0.2 & $L\times L$ & 28--63 & \textemdash{} & 30 & 53000--740\footnotemark[1]\footnotetext{$M$ depends on $L$~\cite{SM}.}
			& \textemdash{} & 2 & $-3.6$ & 0.2 & $L\times L$ & 35--63 & \textemdash{} & \textemdash{} & 25--31\footnotemark[2]\footnotetext{$N_{g}=25$ for $L\le 49$ and $N_{g}=31$ for $L=56,63$.} & 32000--290\footnotemark[1] & \textemdash{}\tabularnewline
			II & 0.2 & $L\times L$ & 21--77 & \textemdash{} & \textemdash{} & 30000 & \textemdash{} & 2 & $-3.6$ & 0.2 & $L\times L$ & 28--77 & \textemdash{} & \textemdash{} & \textemdash{} & 3000--1500\footnotemark[3]\footnotetext{$M=3000$ for $L \le 63$ and $M=1500$ for $L=70,77$.} & \textemdash{}\tabularnewline
			III & 0.2 & $L\times L_{x}$ & 14--210 & $2\cdot10^{7}$ & \textemdash{} & 5 & $10^{8}$ & 5.5 & $-3.6$ & 0.2 & $L\times L_{x}$ & 14--49 & 105 & 5 & \textemdash{} & 15000 & $1.3\cdot10^{6}$\tabularnewline
		\end{tabular}
	\end{ruledtabular}
	
	\caption{\label{tab:parameters} Methods parameters: $W$ is the disorder strength, $L$ and $L_{x}$ the system
		size in the $y$ and $x$ directions, respectively, $N_{g}$ the grid size (method I), $M$ the number of disorder realizations, $L_{x}^{\mathrm{eff}}$
		the effective $x$-length (method III), $p$ the
		hopping range cutoff (method III, nonequilibrium), $\mu^{\mathrm{eff}}$ the
		effective chemical potential, and $\gamma^{\mathrm{in}}/\gamma^{0}$
		the refilling rate in units of $\gamma^{0} = 2\pi\nu_{0}t^2$.}
	
\end{table*}

\textit{Recipe.\textemdash }
We now briefly recall the recipe for the dissipative creation 
of topological states, which is comprehensively described
in Ref.~\cite{Goldstein2019}.
Suppose we have a ``reference Hamiltonian'', $H^{\mathrm{ref}}=\sum_{i,j}h_{ij}^{\mathrm{ref}}c_{i}^{\dagger}c_{j}=
\sum_{\lambda}\varepsilon_{\lambda}^{\mathrm{ref}}c_{\lambda}^{\dagger}c_{\lambda}$ ($i,j$ being real space indexes in 2D, and $\lambda$ an eigenvalue index, which, in the clean case, would correspond to the band number and lattice momentum),
with some desired (e.g., topologically-nontrivial) gapped ground state where only low-lying states ($\lambda \le \lambda_0$) are filled.
Rather than implementing $H^{\mathrm{ref}}$ as the system Hamiltonian,
one may set the system Hamiltonian to zero and employ dissipation to drive the system into a
steady-state which is close to the ground state of $H^{\mathrm{ref}}$.
For this one takes a system consisting of two types of fermions (e.g., cold atom hyperfine states), with respective creation operators $a^\dagger_i$ (system) and $b^\dagger_i$ (bath).
Both fermion species
feel a lattice potential in the $xy$ plane, but the bath $b$-fermions could also escape in the $z$ direction.
Besides that, the Hamiltonian of the $a$-fermions is trivial, ideally featuring no hopping; deviations from this will be described by a system Hamiltonian $H_{S}=\sum_{i,j} h_{S,ij} a_{i}^{\dagger} a_{j}$.
Rather, the dynamics originates from the system-bath coupling Hamiltonian, which is built out of the matrix elements of the reference Hamiltonian. \red{In the rotating frame (with respect to the system and bath Hamiltonians) it acquires a time-independent form,}
\begin{align}
H_{SB} & = \sum_{i,j} \left( h_{ij}^{\mathrm{ref}} -\mu^{\mathrm{eff}} \delta_{ij} \right) b_{i}^{\dagger}a_{j} + \mathrm{h.c.}
\nonumber \\
& = 
\sum_{\lambda}\left(\varepsilon_{\lambda}^{\mathrm{ref}}-\mu^{\mathrm{eff}}\right)b_{\lambda}^{\dagger}a_{\lambda}+\mathrm{h.c.},
\end{align}
\red{where $\mu^{\mathrm{eff}}$ is an effective ``chemical potential''.} The utility of the construction now becomes apparent: 
Suppose the lowest \red{energy} band of the reference Hamiltonian is almost flat (dispersionless). By tuning $\mu^{\mathrm{eff}}$ to its center ($\varepsilon_{\lambda}^{\mathrm{ref}}\approx\mu^{\mathrm{eff}}$ for all $\lambda \le \lambda_0$),
its states becomes \textit{weakly coupled} to the bath compared to
states in the other bands, $\lambda > \lambda_0$.
Thus, all states are evaporated rapidly, except those belonging to the lowest band.
One may then introduce another similar reservoir which refills all trapped states at a uniform rate. 
\red{Coupling the system to these two reservoirs with different chemical potentials
stabilizes a nonequilibrium} steady state close to the ground state of the reference Hamiltonian, as we now explain.

Integrating out the baths one gets a Lindblad~\cite{CrispinGardiner2004} 
master equation,
from which the Gaussian steady-state $\rho$ can be obtained. The latter is completely characterized by the single-particle density matrix $G_{ij}\equiv\mathrm{tr}(\rho a_{i}^{\dagger}a_{j})$, which
obeys a continuous Lyapunov equation~\cite{Schwarz2016,Goldstein2019,Shavit2020}:
\begin{equation}
i\left[ G, h_{S}^{*} \right] + \frac{1}{2}
\left\{ G, \gamma^{\mathrm{out}}+\gamma^{\mathrm{in}} \right\}
=\gamma^{\mathrm{in}},
\label{eq:continuous_lyapunov}
\end{equation}
where $\gamma^{\mathrm{in}},\gamma^{\mathrm{out}}$ are nonnegative
Hermitian matrices that describes the rates which particles enter/escape
of the system, and $h_{S}^{*}$ is the complex conjugate of the matrix
$h_{S}$. By Fermi's golden rule,
$\gamma_{\lambda}^{\mathrm{out}}=2\pi \nu_{0} (\varepsilon_{\lambda}^{\mathrm{ref}}-\mu^{\mathrm{eff}} )^{2}$
is diagonal in the eigenbasis of the reference Hamiltonian [more generally, as a matrix $\gamma^{\mathrm{out}}=2\pi \nu_{0} (h^\mathrm{ref}-\mu^{\mathrm{eff}}\mathbb{I})^{2}$],
with $\nu_{0}$ the density of states of the $b$-species (assumed constant), while
$\gamma^{\mathrm{in}}$
is taken as state independent (proportional to the unit matrix). 
For $h_S=0$,
\newred{we can solve Eq. (\ref{eq:continuous_lyapunov}) explicitly:
\begin{equation}
	G=[1+\frac{2\pi\nu_0}{\gamma^{\mathrm{in}}}(h^{\mathrm{ref}}-\mu^{\mathrm{eff}}\mathbb{I})^{2}]^{-1}.
	\label{eq:G_for_h_zero}
\end{equation}
We see that} $G$ is diagonal in the eigenbasis of $H^\mathrm{ref}$, with eigenvalues
$n_{\lambda} = \gamma^{\mathrm{in}} / (\gamma^{\mathrm{in}} + \gamma_{\lambda}^\mathrm{out})$
representing their mean occupation.
\newred{The coupling to two reservoirs with different chemical potentials thus induced a Lorentzian nonequilibrium distribution (in terms of the energies of $H^\mathrm{ref}$), unlike the equilibrium Fermi-Dirac distribution.}
$G$ [or, equivalently, the system-bath entanglement Hamiltonian $-\ln(\rho)$] has a similar band structure to $H^\mathrm{ref}$ \red{(with the highest occupancy band of $G$ corresponding to the lowest energy band of $H^\mathrm{ref}$)}, which is amenable to topological classification~\cite{Diehl2011,Bardyn2012,Bardyn2013,Budich2015,Budich2015b,Goldstein2019,Shavit2020}.
For $\max_{\lambda\le\lambda_0} (\red{\gamma^\mathrm{out}_{\lambda}}) \ll \gamma^\mathrm{in} \ll \min_{\lambda>\lambda_0} (\red{\gamma^\mathrm{out}_{\lambda}})$ we get $n_{\lambda\le\lambda_0} \approx 1$, $n_{\lambda>\lambda_0} \approx 0$, as desired:
\newred{The steady-state is then close to the ground state of $H^{\mathrm{ref}}$ at zero temperature, and will therefore have the same topological index.}
This motivates the study of \newred{the eigenmodes of $G$ and their localization properties} in the presence of disorder.


\textit{Localization transition.\textemdash} This work compares the localization quantum phase transition of two systems. The first is the equilibrium Hofstadter model~\cite{Hofstadter1976} for the integer
quantum Hall effect on a square lattice:
\begin{equation}
\negthickspace H_H {=} t \negthickspace \sum_{r_{x},r_{y}} \negthickspace e^{2\pi i\alpha r_{y}}a_{r_{x}{+}1,r_{y}}^{\dagger}a_{r_{x},r_{y}}
{+}
a_{r_{x},r_{y}{+}1}^{\dagger}a_{r_{x},r_{y}}
{+}
\mathrm{h.c.},
\label{eq:Hofstadter}
\end{equation}
where we take $t=1$, $\alpha=1/7$. The second system is the out of
equilibrium analog, built using the recipe described above~\cite{Goldstein2019,Shavit2020}: The
Hofstadter Hamiltonian (whose lowest band is naturally almost-flat) is taken as the \textit{reference} Hamiltonian
$H^{\mathrm{ref}}=H_H$, while $H_{S}=0$. To study the localization
phase transition we introduce disorder. In equilibrium
we add a term $H_D = \sum_{r_{x},r_{y}} w_{r_{x},r_{y}} a_{r_{x},r_{y}}^{\dagger} a_{r_{x},r_{y}}$,
where $w_{r_{x},r_{y}} \in [-W,W]$ are independent and uniformly distributed. Out of equilibrium, there are two options for introducing the same disorder term,
\red{realizable in cold atoms using the setup introduced in Refs.~\cite{Goldstein2019,Shavit2020}: One may either
(a) add $H_D$ to $H^{\mathrm{ref}}$ while keeping $H_S=0$, by adding a random component to the laser beam which drives the onsite $a \to b$ transition in $H_{SB}$, using, e.g., a speckle pattern ~\cite{Goodman2020}; (b) keep $H^\mathrm{ref}=H_H$
and set $H_{S}=H_D$, by adding a random component to the optical lattice potential of the $a$ atoms or to the optical potential confining them to the lattice plane.}
We find that in both cases the disorder causes a nonequilibrium \red{steady-state} localization phase transition of the eigenmodes of $G$.
\red{
We can define the localization length of an eigenmode of $G$ by the exponential decay of its envelope, in the same way it is defined for the eigenmodes of $H$ in equilibrium~\cite{Huckestein1995,Evers2008}. Similarly to Eq.~(\ref{eq:localization_length}), it behaves as $\xi(n)\propto\left|n-n_{c}\right|^{-\nu}$, where now it depends on the eigenvalue of $G$, that is, the occupation $n$ (instead of the energy $E$). $n_{c}$ is the critical occupation, which replaces the critical energy $E_c$.
In this work we will concentrate on the band of highest occupation, akin to the lowest Landau band in equilibrium [see for example the bottom panel of Fig.~\ref{fig:local_Chern_number}(b)].}

Does $\nu$ takes the same value as in equilibrium? In the first case
the answer is yes; since $H_{S}=0$,
\newred{$G$ is still given by Eq.~(\ref{eq:G_for_h_zero}).}
Thus, even in the presence of disorder, $h^{\mathrm{ref}}$ and $G$
share the \textit{same eigenvectors}, hence the same $\nu$~\cite{SM}. 
This argument does not hold in the second scenario (disorder in $H_{S}$),
since $G$ and $h^{\mathrm{ref}}$ have different eigenvectors.
Here we need to resort to numerical solution of Eq.~(\ref{eq:continuous_lyapunov}). We will investigate $\nu$ using three FSS
methods. For each we first calculate $\nu$ in equilibrium (disordered Hofstadter
model), and then out of equilibrium ($H^{\mathrm{ref}} = H_H$
and $H_{S}=H_D$).
Again, while in equilibrium we examine the properties of the Hamiltonian (e.g.,
eigenvector localization length, Chern number), out of equilibrium we investigate \red{the same properties, which are now obtained from $G$ instead of the Hamiltonian}.
\red{While the band structure in equilibrium depends only on $\alpha$ and the disorder strength $W$, out of equilibrium it also depends on $\gamma^{\mathrm{in}}$ and $\mu^{\mathrm{eff}}$. The results were found not to be sensitive to their particular values, as long as they are chosen so that the disorder broadens the bands more than their clean width but less than their separation~\cite{SM}.}
The parameter values are summarized in Table~\ref{tab:parameters}, and the final
results in Table~\ref{tab:nu_results}.

\begin{figure}
	\includegraphics[scale=0.32]{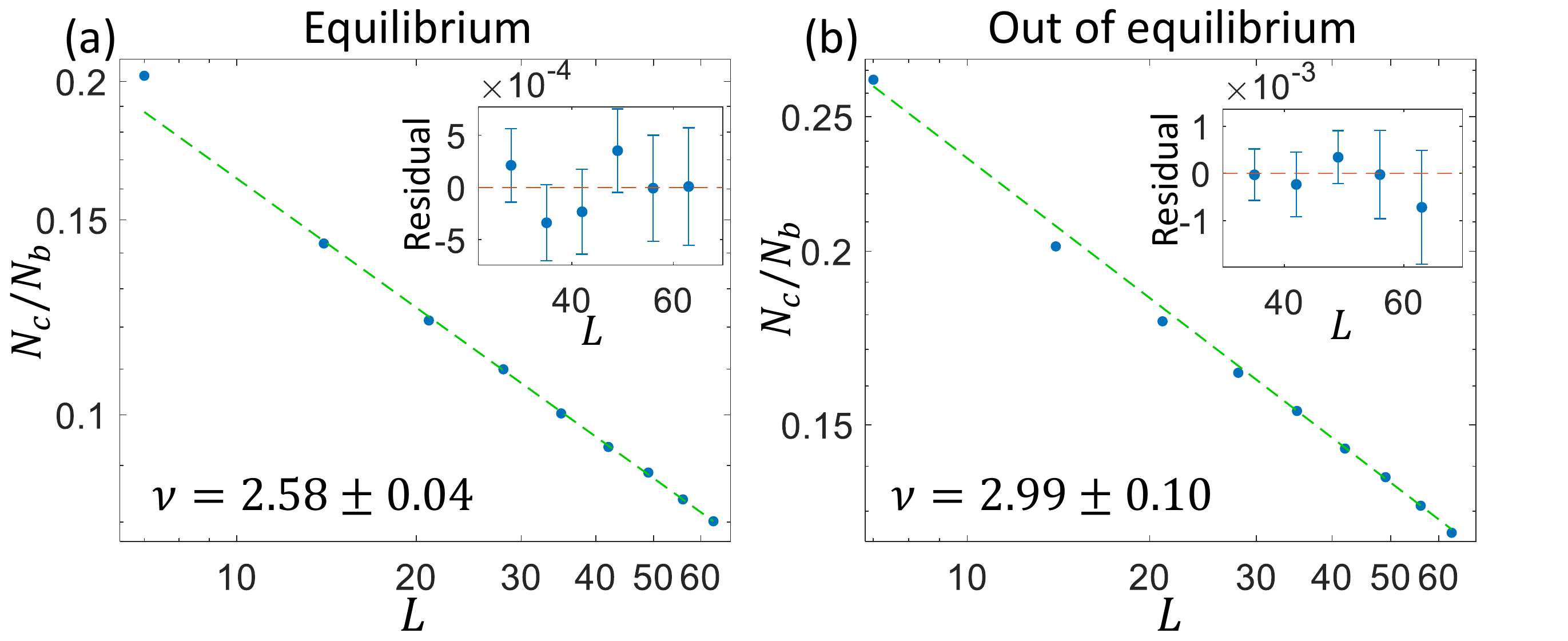}
	\caption{\label{fig:coundcting_states_results}
		Log-log plot of $\left\langle N_{c}/N_{b}\right\rangle $
		as function of $L$ (system size), with $N_{c}$ the number of conducting
		states and $N_{b}=\alpha L^{2}$ ($\alpha=1/7$) the total number of states per band.
		Dashed lines represent linear fits with $L\ge28$ in equilibrium
		and $L\ge35$ out of equilibrium. Insets: residual plots.}
\end{figure}

\textit{Method I.\textemdash }
Following Ref.~\cite{Zhu2019,SM}, we calculate the critical
exponent in equilibrium by the scaling of the number of conducting states,
$N_{c}$, 
\begin{equation}
N_{c}(L)\propto L^{2-1/\nu},\label{eq:conducting_states}
\end{equation}
where $L$ is the system size and $\nu$ is the critical exponent.
Working with a $L\times L$ Hofstadter model with periodic boundary conditions,
we calculate $N_{c}$ by counting the number of single-particle states
with nonzero Chern number, and average the result over $M$ different
disorder realizations. In the presence of disorder, the Chern
number can be defined as~\cite{Niu1985}:
\begin{equation}
C_{L}(\psi)=-\dfrac{1}{\pi}\int \mathrm{Im}\left\langle \partial_{\theta_{x}}\psi|\partial_{\theta_{y}}\psi\right\rangle d\theta_{x}d\theta_{y},\label{eq:flux_chern_number}
\end{equation}
where $\psi(\theta_{x},\theta_{y})$ is the single-particle state and the integral is over
the space of twisted periodic boundary conditions, defined by the phases $0\le\theta_{x},\theta_{y}\le2\pi$.
For efficient calculation, we use the method suggested in Ref.~\cite{Fukui2005}, employing grid size $N_{g}\times N_{g}$~\cite{SM}.
Corrections to the scaling in Eq.~(\ref{eq:conducting_states}) 
fade quickly with increasing the system size, hence may be ignored
by excluding low system sizes. The nonequilibrium
generalization is straight-forward: We calculate the Chern number
of eigenstates of $G$ (instead of $H$) by introducing the twisted boundary
conditions $\theta_{x},\theta_{y}$ 
into $H^\mathrm{ref}$. 
\red{Then, we count the conducting states within the highest occupation band.}
Results are presented in Fig.~\ref{fig:coundcting_states_results}.

\begin{figure}
\includegraphics[scale=0.32]{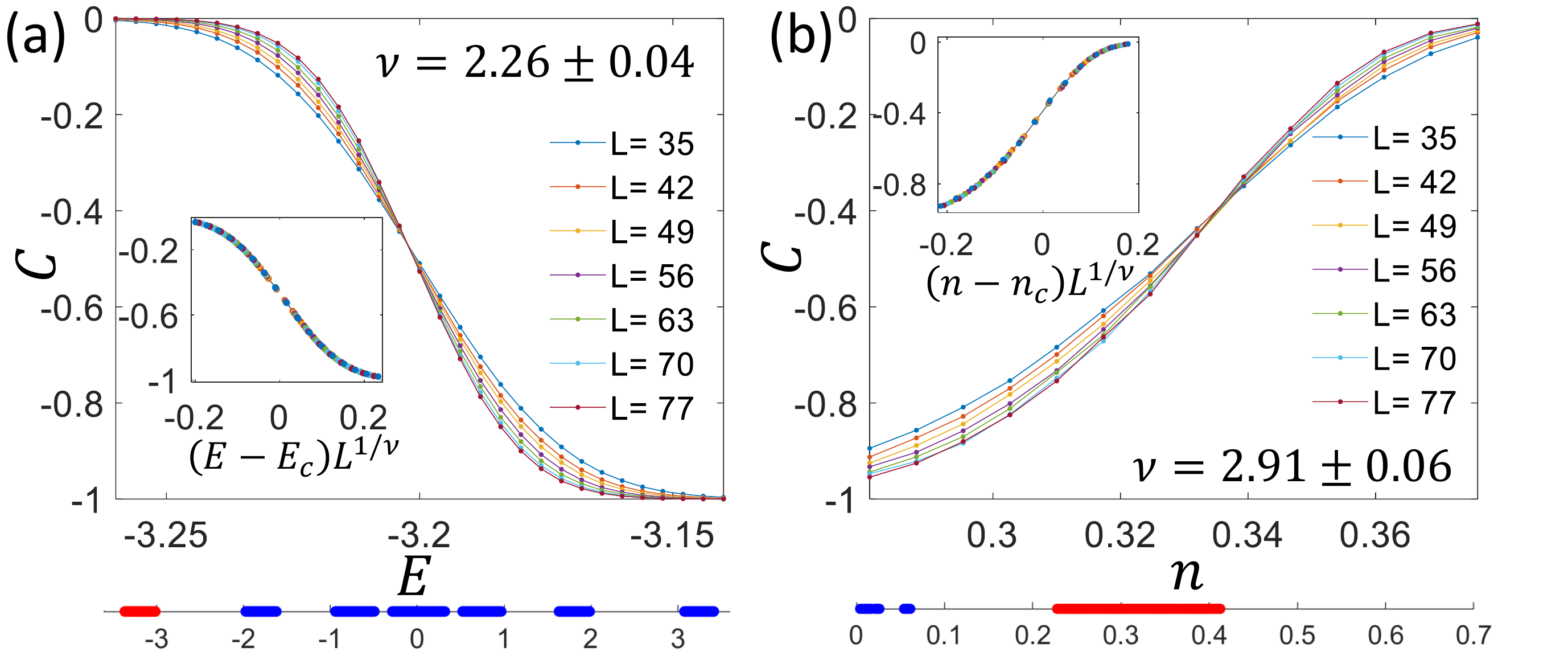}
	\caption{\label{fig:local_Chern_number} The average local Chern number (a) in and (b) out of equilibrium. Insets: scaling data collapse. \red{Bottom panels: the seven
		energy bands in equilibrium, and occupation bands out of equilibrium (note the different scales). The band which is investigated is
		marked in red and the others in blue.}
	}
\end{figure}

\textit{Method II.\textemdash }
Here we study FSS of the topological index~\cite{Huckestein1995,SM}.
\red{In equilibrium, we define the total Chern number $C_{L}(E)$
as the sum of the Chern numbers defined in Eq.~(\ref{eq:flux_chern_number}) over all single particle states $\psi$ with energy below $E$ (hence it varies between 0 when $E$ is below the lowest band, to $-1$ when it is in the gap between it and the next band).
In the vicinity of the critical energy $E_{c}$, it scales as 
\begin{equation}
C_{L}(E)=f\left((E-E_{c})L^{1/\nu}\right),
\label{eq:chern_number_scaling}
\end{equation}
We note that the transition will be sharp in the thermodynamic limit.
For a more efficient estimation of $C_{L}(E)$, we will use}
the local Chern marker \cite{Bianco2011,Caio2019} with open boundary conditions, 
\begin{equation}
C(r_{x},r_{y})=-2\pi i\left\langle r_{x},r_{y} \right|\tilde{X}\tilde{Y}-\tilde{Y}\tilde{X}\left| r_{x},r_{y} \right\rangle,
\label{eq:local_chern_number}
\end{equation}
where $\tilde{X},\tilde{Y}$ are the projected lattice position operators:
\red{$\tilde{X}=P(E) X P(E),\quad\tilde{Y}=P(E) Y P(E)$, $P(E)$ being a projection onto states with energy below $E$.}
The local Chern marker
fluctuates around the value of the Chern number in the bulk of the system,
but takes different values on the edges, so that $\sum_{r_{x},r_{y}} C(r_{x},r_{y}) = 0$.
Thus, 
we average
$C(r_{x},r_{y})$
over the bulk, while excluding 1/4 of the
sample length from each side, 
$C_{L}(E)=(4/L^2) \times \sum_{L/4 \le r_{x},r_{y} \le 3L/4 }C(r_{x},r_{y})$, and average the result over $M$ different disorder realizations.
As in method I,
irrelevant corrections exist, but their influence decreases rapidly with increasing system size.
We then search for $\nu,\,E_{c}$, and the coefficients 
of a polynomial approximating $f$~\cite{SM}, 
which minimize the \red{chi-squared deviation of $C_L(E)$ from the scaling Eq.~(\ref{eq:chern_number_scaling}).}
Out of equilibrium, we calculate \red{$C_{L}(n)$,
	the Chern number of eigenstates of $G$  with occupation \emph{larger} than $n$, using Eq.~(\ref{eq:local_chern_number}) with the appropriate projector $P(n)$.
	}
The results are presented in Fig.~\ref{fig:local_Chern_number}.

\begin{figure}
	\includegraphics[scale=0.4]{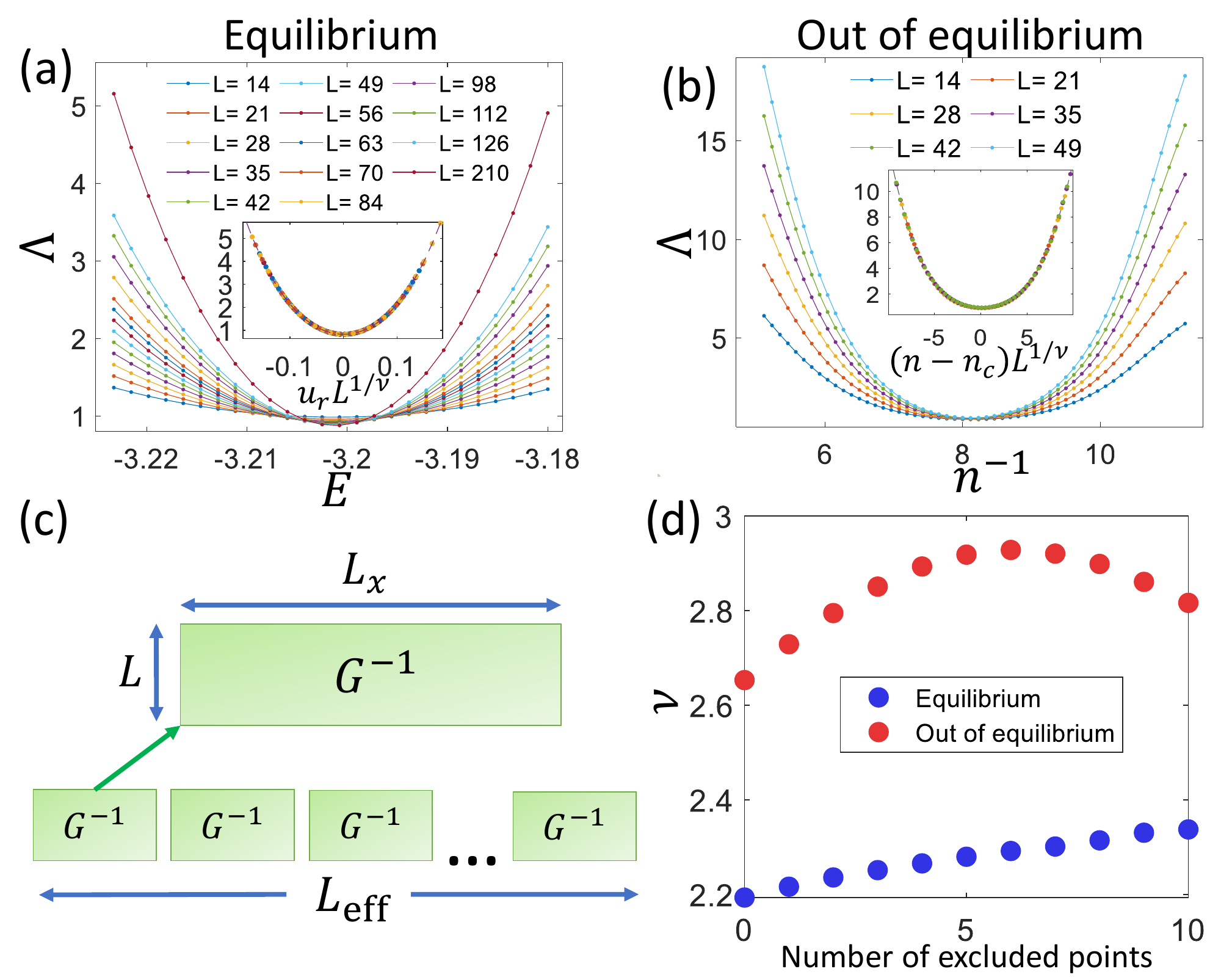}
	\caption{\label{fig:localization_length}  Dimensionless
		Lyapunov exponent (a) in and (b) out of equilibrium. \red{Insets: scaling data collapse \{in (a) the vertical axis includes corrections to scaling and $u_r$ is the relevant scaling field~\cite{SM}\}.} (c) Illustration
		of the nonequilibrium transfer matrices construction.
		(d) Comparison of the critical exponent in and out of equilibrium,
		without corrections to scaling. The horizontal axis represents the number
		data points 
		that were excluded from each side of the critical point in the chi-squared minimization.
	}
\end{figure}

\textit{Method III.\textemdash}
Here we perform FSS of the localization length $\xi$. Following Ref.~\cite{MacKinnon1983} (see also~\cite{SM}),
we calculate the localization length with the transfer-matrix method:
We consider a long cylinder of size $L_{x}\times L$, $L_{x} \gg L$.
Let $\psi$ be an eigenvalue of the Hamiltonian with energy $E$.
From the equation $H\psi=E\psi$ we can construct the $2L\times2L$ transfer-matrix
$T_{r_{x}}$, defined as: 
\begin{equation}
\left(\begin{array}{c}
\psi_{r_{x}+1}\\
\psi_{r_{x}}
\end{array}\right)=T_{r_{x}}\left(\begin{array}{c}
\psi_{r_{x}}\\
\psi_{r_{x}-1}
\end{array}\right),\label{eq:transfer_matrix}
\end{equation}
where $\psi_{r_{x}}$ is a vector with $L$ elements $\psi_{r_{x},r_{y}=1 \cdots L}$. 
Being symplectic, the eigenvalues of each transfer matrix come in
reciprocal pairs $\{ \lambda,\lambda^{-1} \}$. The same applies to their
product, $\mathcal{T}=\prod_{r_x=1}^{L_{x}}T_{r_x}$. The Lyapunov exponent (inverse localization length)
is defined as:
\begin{equation}
\tilde{\Lambda} \equiv \xi^{-1} = \lim_{L_{x}\rightarrow\infty}\frac{\ln(\lambda_\mathrm{min})}{L_{x}},\label{eq:Lyapunov_exponent}
\end{equation}
where $\lambda_\mathrm{min}$ is the smallest eigenvalue of $\mathcal{T}$ that is
larger than unity. We have applied the Gram-Schmidt process to the columns
of $\mathcal{T}$ every $7$ multiplications to reduce numerical error.
The results are presented in Fig.~\ref{fig:localization_length}(a).
As in the previous method, $\nu$ can be extracted by 
	\red{finding a function $f$ that minimize the chi-square of the} dimensionless Lyapunov exponent
	$\Lambda\equiv L\tilde{\Lambda}$.
However, since the data
contains strong corrections to scaling
(
typical for the
long cylinder geometry), we account for a single irrelevant
scaling field~\cite{SM}.

The nonequilibrium generalization from $\Lambda(E)$ to $\Lambda(n)$ is more complicated compared
to the previous methods. 
First, unlike $H$, $G$ has \textit{non-local hopping terms} which prevent
us from constructing a transfer matrix. This requires introducing 
a cutoff $p$ on the hopping range in the $x$ direction, and setting terms of range larger
than $p$ to zero. 
From this perspective it is advantageous to construct the transfer matrices using $G^{-1}$, since Eq.~(\ref{eq:G_for_h_zero}) shows that for $H_S=0$ its elements have a finite range $p=2$. We have verified numerically that the elements of $G^{-1}$ decay exponentially with range for $H_S=H_D$, making truncation at $p=5$ a very good approximation~\cite{SM}.

A second issue is that in the presence of disorder the structure of $G^{-1}$ can only be obtained numerically, by solving Eq.~(\ref{eq:continuous_lyapunov}). 
Thus, we cannot analytically obtain the transfer matrix at a specific $x$-position, and
instead, we can only generate the entire $G^{-1}$ matrix, which is
impractical for $L_{x}\gg1$. As a solution,
we use the scheme depicted in Fig.~\ref{fig:localization_length}(c): We generate a $G^{-1}$ matrix of size $L_{x}\times L$ for some
large but practical $L_{x}$ (with periodic boundary conditions)~\cite{SM}. 
We repeat this with $M$ disorder realizations, and denote the resulting matrices
as $\left\{ \left(G^{-1}\right)_{m}\right\} _{m=1}^{M}$. From each
$\left(G^{-1}\right)_{m}$ we extract $K$ transfer matrices ($K = L_{x}-2c$, excluding the $c=7$ matrices closest to each end) by imposing a cutoff $p$ on the hopping range, as explained
above. 
We then define the sequence $\left\{ T_{n}\right\} _{n=1}^{MK}$,
with $T_{(n-1)K+1},...,T_{nK}$ the transfer matrices extracted
from $\left(G^{-1}\right)_{n}$. 
The effective system length is thus $L_{\mathrm{eff}}=MK$. 
The mismatch
between transfer-matrices that originate from different $G^{-1}$ (for
example, $T_{K}$ and $T_{K+1}$) introduces an error, but it
can be reduced by increasing $L_{x}$~\cite{SM}.

The results are shown in Fig.~\ref{fig:localization_length}(b). The numerical effort per sample is still much higher in the
nonequilibrium case, limiting our ability to reduce statistical
error by either sample averaging or using large system sizes. Hence, we can neither implement corrections to scaling nor
drop small systems, and therefore cannot determine $\nu$ as accurately
as before. We thus resort to extracting the uncorrected nonequilibrium
exponent and comparing it with a similarly obtained equilibrium
value, to appreciate the significance of their difference, see Fig.~\ref{fig:localization_length}(d).

\begin{table}
	\begin{ruledtabular}
		\begin{tabular}{c|ccc}
			Method & I & II & III \\
			\hline
			Equilibrium & $2.58 \pm 0.04$ & $2.26 \pm 0.04$ & $2.53 \pm 0.03$ \\
			Nonequilibrium &  $2.99 \pm 0.10$ & $2.91 \pm 0.06$ &
			\multirow{2}{1.5in}{not convergent, higher than equilibrium} \\
			\\
		\end{tabular}
	\end{ruledtabular}
	\caption{\label{tab:nu_results} Summary of the results for the critical exponent
		$\nu$, in and out of equilibrium.}
\end{table}

\textit{Results and Discussion.\textemdash}
The results are summarized in Table~\ref{tab:nu_results}. In equilibrium they are generally in line with previous studies
\red{~\cite{Slevin2009,Obuse2010,Amado2011,Fulga2011,Slevin2012,Obuse2012,Nuding2015,Gruzberg2017,Ippoliti2018,Puschmann2019,Zhu2019,Sbierski2021}}. 
For method I, the obtained $\nu=2.58\pm0.04$ is somewhat higher than the value $\nu=2.50\pm0.01$ reported in Ref.~\cite{Zhu2019}
(also for $\alpha=1/7$). This might be related to the fact that there
the disorder Hamiltonian has been projected to the clean lowest band. 
In method II, the result ($\nu=2.26\pm0.04$) is 
smaller than recent estimates of the critical exponent, which seems to be a general feature of FSS of a topological index~\cite{Loring2010,Ippoliti2018}. \red{Let us note that in any case we are interested in the equilibrium-nonequilibrium difference, which is larger than this discrepancy.}
In method III,
upon including corrections to scaling we get $\nu=2.53\pm0.03$, $y=0.44\pm0.01$, $\Lambda(E_{c})=0.83\pm0.01$,
with $y$ the leading irrelevant exponent. 
This is slightly smaller but still in agreement with $\nu=2.58\pm0.03$
obtained in Ref.~\cite{Puschmann2019} for the Hofstadter
model.

Out of equilibrium, methods I and II give rise to values which are significantly
higher than in equilibrium. The results of method III are not convergent, 
but they still strongly suggest that $\nu$ is higher than equilibrium by 0.5--0.6 \red{(see. Fig.~\ref{fig:localization_length}(d))}, in agreement with the other
methods. 
\newred{We have also verified that our results are insensitive} to the specific parameter values~\cite{SM}. \newred{All this points at a different type of nonequilibrium universality class.}

Let us reiterate that the single-particle density matrix $G$ is Hermitian. Furthermore, $G^{-1}$ is local in space. The locality is exact for $h_S=0$, where $G^{-1}$ is essentially the square of $h^\mathrm{ref}$, see Eq.~(\ref{eq:G_for_h_zero}). 
We have found that for disorder in $H_S$ the elements of $G^{-1}$ have distributions without fat tails, and with averages and correlations which decay exponentially with distance~\cite{SM}. Thus, our results indicate a different type of universality class of the local Hermitian disordered $G^{-1}$, which is rooted in the nonequilibrium nature of the system.

\red{The value of $\nu$ could be measured experimentally, by using the following protocol: (i) realize the cold atoms setup described in Ref.~\cite{Goldstein2019}; (ii) use  a laser speckle~\cite{Goodman2020} to introduce disorder, either in the beam that induce the $a\rightarrow b$ transitions (for disorder in $H^{\mathrm{ref}}$), or in the beam that is responsible for the confinement of the $a$-atoms (for disorder in $H_{S}$), as discussed above; (iii) measure $G$ as demonstrated in Refs.~\cite{Hauke2014,Flaschner2016,Tarnowski2017,Ardilla2018,Zheng2020}; and (iv) repeat for different system sizes to extract $\nu$ through FSS.}

\textit{Conclusions.\textemdash}
In this work we have investigated the effects of disorder on dissipation-induced topological states.
We demonstrated the existence of \red{nonequilibrium steady-state} localization phase transition
similar to the integer quantum Hall plateau transition. Using
three FSS methods, we found a significant difference between the value of the
critical exponent $\nu$ in and out of equilibrium when disorder is introduced into the non-dissipative part of the Lindbladian. This indicates a different type of nonequilibrium quantum universality class, despite the steady state density matrix being Hermitian and local.
\red{Our findings could be tested in cold-atom experiments.} 
In the future it would be interesting \red{to investigate other types of disorder (e.g., long range~\cite{Fogler1998,Ostrovsky2007,Rycerz2007})},  to attack the problem using field theoretical methods~\cite{Evers2008,Sieberer2016}, and to study the relation between the steady state and the nonhermitian~\cite{Hatano1996,Ashida2020,Silberstein2020} decay towards it (a relation which is nontrivial out of equilibrium~\cite{Shavit2020}), as well as the possibility of new many-body localization transition~\cite{Nandkishore2015,Altman2015}.

\begin{acknowledgments}
We thank I.S.~Burmistrov, R.~Ilan, and E.~Shimshoni for useful discussions.
Support by the Israel Science Foundation (Grant No. 227/15) and the US-Israel Binational Science Foundation (Grant No. 2016224) is gratefully acknowledged.
\end{acknowledgments}

\bibliography{bibi}

\begin{thebibliography}{77}%
\makeatletter
\providecommand \@ifxundefined [1]{%
 \@ifx{#1\undefined}
}%
\providecommand \@ifnum [1]{%
 \ifnum #1\expandafter \@firstoftwo
 \else \expandafter \@secondoftwo
 \fi
}%
\providecommand \@ifx [1]{%
 \ifx #1\expandafter \@firstoftwo
 \else \expandafter \@secondoftwo
 \fi
}%
\providecommand \natexlab [1]{#1}%
\providecommand \enquote  [1]{``#1''}%
\providecommand \bibnamefont  [1]{#1}%
\providecommand \bibfnamefont [1]{#1}%
\providecommand \citenamefont [1]{#1}%
\providecommand \href@noop [0]{\@secondoftwo}%
\providecommand \href [0]{\begingroup \@sanitize@url \@href}%
\providecommand \@href[1]{\@@startlink{#1}\@@href}%
\providecommand \@@href[1]{\endgroup#1\@@endlink}%
\providecommand \@sanitize@url [0]{\catcode `\\12\catcode `\$12\catcode
  `\&12\catcode `\#12\catcode `\^12\catcode `\_12\catcode `\%12\relax}%
\providecommand \@@startlink[1]{}%
\providecommand \@@endlink[0]{}%
\providecommand \url  [0]{\begingroup\@sanitize@url \@url }%
\providecommand \@url [1]{\endgroup\@href {#1}{\urlprefix }}%
\providecommand \urlprefix  [0]{URL }%
\providecommand \Eprint [0]{\href }%
\providecommand \doibase [0]{https://doi.org/}%
\providecommand \selectlanguage [0]{\@gobble}%
\providecommand \bibinfo  [0]{\@secondoftwo}%
\providecommand \bibfield  [0]{\@secondoftwo}%
\providecommand \translation [1]{[#1]}%
\providecommand \BibitemOpen [0]{}%
\providecommand \bibitemStop [0]{}%
\providecommand \bibitemNoStop [0]{.\EOS\space}%
\providecommand \EOS [0]{\spacefactor3000\relax}%
\providecommand \BibitemShut  [1]{\csname bibitem#1\endcsname}%
\let\auto@bib@innerbib\@empty
\bibitem [{\citenamefont {Kamenev}(2009)}]{Kamenev2009}%
  \BibitemOpen
  \bibfield  {author} {\bibinfo {author} {\bibfnamefont {A.}~\bibnamefont
  {Kamenev}},\ }\href {https://www.doi.org/10.1017/cbo9781139003667} {\emph
  {\bibinfo {title} {Field Theory of Non-Equilibrium Systems}}}\ (\bibinfo
  {publisher} {Cambridge University Press},\ \bibinfo {year}
  {2009})\BibitemShut {NoStop}%
\bibitem [{\citenamefont {Sieberer}\ \emph {et~al.}(2016)\citenamefont
  {Sieberer}, \citenamefont {Buchhold},\ and\ \citenamefont
  {Diehl}}]{Sieberer2016}%
  \BibitemOpen
  \bibfield  {author} {\bibinfo {author} {\bibfnamefont {L.~M.}\ \bibnamefont
  {Sieberer}}, \bibinfo {author} {\bibfnamefont {M.}~\bibnamefont {Buchhold}},\
  and\ \bibinfo {author} {\bibfnamefont {S.}~\bibnamefont {Diehl}},\ }\bibfield
   {title} {\bibinfo {title} {Keldysh field theory for driven open quantum
  systems},\ }\href {https://doi.org/10.1088/0034-4885/79/9/096001} {\bibfield
  {journal} {\bibinfo  {journal} {Reports on Progress in Physics}\ }\textbf
  {\bibinfo {volume} {79}},\ \bibinfo {pages} {096001} (\bibinfo {year}
  {2016})}\BibitemShut {NoStop}%
\bibitem [{\citenamefont {Diehl}\ \emph {et~al.}(2008)\citenamefont {Diehl},
  \citenamefont {Micheli}, \citenamefont {Kantian}, \citenamefont {Kraus},
  \citenamefont {Buchler},\ and\ \citenamefont {Zoller}}]{Diehl2008}%
  \BibitemOpen
  \bibfield  {author} {\bibinfo {author} {\bibfnamefont {S.}~\bibnamefont
  {Diehl}}, \bibinfo {author} {\bibfnamefont {A.}~\bibnamefont {Micheli}},
  \bibinfo {author} {\bibfnamefont {A.}~\bibnamefont {Kantian}}, \bibinfo
  {author} {\bibfnamefont {B.}~\bibnamefont {Kraus}}, \bibinfo {author}
  {\bibfnamefont {H.~P.}\ \bibnamefont {Buchler}},\ and\ \bibinfo {author}
  {\bibfnamefont {P.}~\bibnamefont {Zoller}},\ }\bibfield  {title} {\bibinfo
  {title} {Quantum states and phases in driven open quantum systems with cold
  atoms},\ }\href {https://www.doi.org/10.1038/nphys1073} {\bibfield  {journal}
  {\bibinfo  {journal} {Nature Physics}\ }\textbf {\bibinfo {volume} {4}},\
  \bibinfo {pages} {878} (\bibinfo {year} {2008})}\BibitemShut {NoStop}%
\bibitem [{\citenamefont {Kraus}\ \emph {et~al.}(2008)\citenamefont {Kraus},
  \citenamefont {Buchler}, \citenamefont {Diehl}, \citenamefont {Kantian},
  \citenamefont {Micheli},\ and\ \citenamefont {Zoller}}]{Kraus2008}%
  \BibitemOpen
  \bibfield  {author} {\bibinfo {author} {\bibfnamefont {B.}~\bibnamefont
  {Kraus}}, \bibinfo {author} {\bibfnamefont {H.~P.}\ \bibnamefont {Buchler}},
  \bibinfo {author} {\bibfnamefont {S.}~\bibnamefont {Diehl}}, \bibinfo
  {author} {\bibfnamefont {A.}~\bibnamefont {Kantian}}, \bibinfo {author}
  {\bibfnamefont {A.}~\bibnamefont {Micheli}},\ and\ \bibinfo {author}
  {\bibfnamefont {P.}~\bibnamefont {Zoller}},\ }\bibfield  {title} {\bibinfo
  {title} {Preparation of entangled states by quantum markov processes},\
  }\href {https://www.doi.org/10.1103/physreva.78.042307} {\bibfield  {journal}
  {\bibinfo  {journal} {Physical Review A}\ }\textbf {\bibinfo {volume} {78}},\
  \bibinfo {pages} {042307} (\bibinfo {year} {2008})}\BibitemShut {NoStop}%
\bibitem [{\citenamefont {Verstraete}\ \emph {et~al.}(2009)\citenamefont
  {Verstraete}, \citenamefont {Wolf},\ and\ \citenamefont
  {Cirac}}]{Verstraete2009}%
  \BibitemOpen
  \bibfield  {author} {\bibinfo {author} {\bibfnamefont {F.}~\bibnamefont
  {Verstraete}}, \bibinfo {author} {\bibfnamefont {M.~M.}\ \bibnamefont
  {Wolf}},\ and\ \bibinfo {author} {\bibfnamefont {J.~I.}\ \bibnamefont
  {Cirac}},\ }\bibfield  {title} {\bibinfo {title} {Quantum computation and
  quantum-state engineering driven by dissipation},\ }\href
  {https://www.doi.org/10.1038/nphys1342} {\bibfield  {journal} {\bibinfo
  {journal} {Nature Physics}\ }\textbf {\bibinfo {volume} {5}},\ \bibinfo
  {pages} {633} (\bibinfo {year} {2009})}\BibitemShut {NoStop}%
\bibitem [{\citenamefont {Weimer}\ \emph {et~al.}(2010)\citenamefont {Weimer},
  \citenamefont {Muller}, \citenamefont {Lesanovsky}, \citenamefont {Zoller},\
  and\ \citenamefont {Buchler}}]{Weimer2010}%
  \BibitemOpen
  \bibfield  {author} {\bibinfo {author} {\bibfnamefont {H.}~\bibnamefont
  {Weimer}}, \bibinfo {author} {\bibfnamefont {M.}~\bibnamefont {Muller}},
  \bibinfo {author} {\bibfnamefont {I.}~\bibnamefont {Lesanovsky}}, \bibinfo
  {author} {\bibfnamefont {P.}~\bibnamefont {Zoller}},\ and\ \bibinfo {author}
  {\bibfnamefont {H.~P.}\ \bibnamefont {Buchler}},\ }\bibfield  {title}
  {\bibinfo {title} {A {R}ydberg quantum simulator},\ }\href
  {https://www.doi.org/10.1038/nphys1614} {\bibfield  {journal} {\bibinfo
  {journal} {Nature Physics}\ }\textbf {\bibinfo {volume} {6}},\ \bibinfo
  {pages} {382} (\bibinfo {year} {2010})}\BibitemShut {NoStop}%
\bibitem [{\citenamefont {Otterbach}\ and\ \citenamefont
  {Lemeshko}(2014)}]{Otterbach2014}%
  \BibitemOpen
  \bibfield  {author} {\bibinfo {author} {\bibfnamefont {J.}~\bibnamefont
  {Otterbach}}\ and\ \bibinfo {author} {\bibfnamefont {M.}~\bibnamefont
  {Lemeshko}},\ }\bibfield  {title} {\bibinfo {title} {Dissipative preparation
  of spatial order in rydberg-dressed bose-einstein condensates},\ }\href
  {https://www.doi.org/10.1103/physrevlett.113.070401} {\bibfield  {journal}
  {\bibinfo  {journal} {Physical Review Letters}\ }\textbf {\bibinfo {volume}
  {113}},\ \bibinfo {pages} {070401} (\bibinfo {year} {2014})}\BibitemShut
  {NoStop}%
\bibitem [{\citenamefont {Lang}\ and\ \citenamefont
  {Buchler}(2015)}]{Lang2015}%
  \BibitemOpen
  \bibfield  {author} {\bibinfo {author} {\bibfnamefont {N.}~\bibnamefont
  {Lang}}\ and\ \bibinfo {author} {\bibfnamefont {H.~P.}\ \bibnamefont
  {Buchler}},\ }\bibfield  {title} {\bibinfo {title} {Exploring quantum phases
  by driven dissipation},\ }\href
  {https://www.doi.org/10.1103/physreva.92.012128} {\bibfield  {journal}
  {\bibinfo  {journal} {Physical Review A}\ }\textbf {\bibinfo {volume} {92}},\
  \bibinfo {pages} {012128} (\bibinfo {year} {2015})}\BibitemShut {NoStop}%
\bibitem [{\citenamefont {Zhou}\ \emph {et~al.}(2017)\citenamefont {Zhou},
  \citenamefont {Choi},\ and\ \citenamefont {Lukin}}]{Zhou2017}%
  \BibitemOpen
  \bibfield  {author} {\bibinfo {author} {\bibfnamefont {L.}~\bibnamefont
  {Zhou}}, \bibinfo {author} {\bibfnamefont {S.}~\bibnamefont {Choi}},\ and\
  \bibinfo {author} {\bibfnamefont {M.~D.}\ \bibnamefont {Lukin}},\ }\bibfield
  {title} {\bibinfo {title} {Symmetry-protected dissipative preparation of
  matrix product states},\ }\Eprint {https://arxiv.org/abs/1706.01995}
  {arXiv:1706.01995 [quant-ph]}  (\bibinfo {year} {2017})\BibitemShut {NoStop}%
\bibitem [{\citenamefont {Diehl}\ \emph {et~al.}(2011)\citenamefont {Diehl},
  \citenamefont {Rico}, \citenamefont {Baranov},\ and\ \citenamefont
  {Zoller}}]{Diehl2011}%
  \BibitemOpen
  \bibfield  {author} {\bibinfo {author} {\bibfnamefont {S.}~\bibnamefont
  {Diehl}}, \bibinfo {author} {\bibfnamefont {E.}~\bibnamefont {Rico}},
  \bibinfo {author} {\bibfnamefont {M.~A.}\ \bibnamefont {Baranov}},\ and\
  \bibinfo {author} {\bibfnamefont {P.}~\bibnamefont {Zoller}},\ }\bibfield
  {title} {\bibinfo {title} {Topology by dissipation in atomic quantum wires},\
  }\href {https://www.doi.org/10.1038/nphys2106} {\bibfield  {journal}
  {\bibinfo  {journal} {Nature Physics}\ }\textbf {\bibinfo {volume} {7}},\
  \bibinfo {pages} {971} (\bibinfo {year} {2011})}\BibitemShut {NoStop}%
\bibitem [{\citenamefont {Bardyn}\ \emph {et~al.}(2012)\citenamefont {Bardyn},
  \citenamefont {Baranov}, \citenamefont {Rico}, \citenamefont
  {{\.{I}}mamo{\u{g}}lu}, \citenamefont {Zoller},\ and\ \citenamefont
  {Diehl}}]{Bardyn2012}%
  \BibitemOpen
  \bibfield  {author} {\bibinfo {author} {\bibfnamefont {C.-E.}\ \bibnamefont
  {Bardyn}}, \bibinfo {author} {\bibfnamefont {M.~A.}\ \bibnamefont {Baranov}},
  \bibinfo {author} {\bibfnamefont {E.}~\bibnamefont {Rico}}, \bibinfo {author}
  {\bibfnamefont {A.}~\bibnamefont {{\.{I}}mamo{\u{g}}lu}}, \bibinfo {author}
  {\bibfnamefont {P.}~\bibnamefont {Zoller}},\ and\ \bibinfo {author}
  {\bibfnamefont {S.}~\bibnamefont {Diehl}},\ }\bibfield  {title} {\bibinfo
  {title} {Majorana modes in driven-dissipative atomic superfluids with a zero
  {C}hern number},\ }\href {https://www.doi.org/10.1103/physrevlett.109.130402}
  {\bibfield  {journal} {\bibinfo  {journal} {Physical Review Letters}\
  }\textbf {\bibinfo {volume} {109}},\ \bibinfo {pages} {130402} (\bibinfo
  {year} {2012})}\BibitemShut {NoStop}%
\bibitem [{\citenamefont {Bardyn}\ \emph {et~al.}(2013)\citenamefont {Bardyn},
  \citenamefont {Baranov}, \citenamefont {Kraus}, \citenamefont {Rico},
  \citenamefont {Imamoglu}, \citenamefont {Zoller},\ and\ \citenamefont
  {Diehl}}]{Bardyn2013}%
  \BibitemOpen
  \bibfield  {author} {\bibinfo {author} {\bibfnamefont {C.-E.}\ \bibnamefont
  {Bardyn}}, \bibinfo {author} {\bibfnamefont {M.~A.}\ \bibnamefont {Baranov}},
  \bibinfo {author} {\bibfnamefont {C.~V.}\ \bibnamefont {Kraus}}, \bibinfo
  {author} {\bibfnamefont {E.}~\bibnamefont {Rico}}, \bibinfo {author}
  {\bibfnamefont {A.}~\bibnamefont {Imamoglu}}, \bibinfo {author}
  {\bibfnamefont {P.}~\bibnamefont {Zoller}},\ and\ \bibinfo {author}
  {\bibfnamefont {S.}~\bibnamefont {Diehl}},\ }\bibfield  {title} {\bibinfo
  {title} {Topology by dissipation},\ }\href
  {https://www.doi.org/10.1088/1367-2630/15/8/085001} {\bibfield  {journal}
  {\bibinfo  {journal} {New Journal of Physics}\ }\textbf {\bibinfo {volume}
  {15}},\ \bibinfo {pages} {085001} (\bibinfo {year} {2013})}\BibitemShut
  {NoStop}%
\bibitem [{\citenamefont {Konig}\ and\ \citenamefont
  {Pastawski}(2014)}]{Koenig2014}%
  \BibitemOpen
  \bibfield  {author} {\bibinfo {author} {\bibfnamefont {R.}~\bibnamefont
  {Konig}}\ and\ \bibinfo {author} {\bibfnamefont {F.}~\bibnamefont
  {Pastawski}},\ }\bibfield  {title} {\bibinfo {title} {Generating topological
  order: No speedup by dissipation},\ }\href
  {https://www.doi.org/10.1103/physrevb.90.045101} {\bibfield  {journal}
  {\bibinfo  {journal} {Physical Review B}\ }\textbf {\bibinfo {volume} {90}},\
  \bibinfo {pages} {045101} (\bibinfo {year} {2014})}\BibitemShut {NoStop}%
\bibitem [{\citenamefont {Kapit}\ \emph {et~al.}(2014)\citenamefont {Kapit},
  \citenamefont {Hafezi},\ and\ \citenamefont {Simon}}]{Kapit2014}%
  \BibitemOpen
  \bibfield  {author} {\bibinfo {author} {\bibfnamefont {E.}~\bibnamefont
  {Kapit}}, \bibinfo {author} {\bibfnamefont {M.}~\bibnamefont {Hafezi}},\ and\
  \bibinfo {author} {\bibfnamefont {S.~H.}\ \bibnamefont {Simon}},\ }\bibfield
  {title} {\bibinfo {title} {Induced self-stabilization in fractional quantum
  {H}all states of light},\ }\href
  {https://www.doi.org/10.1103/physrevx.4.031039} {\bibfield  {journal}
  {\bibinfo  {journal} {Physical Review X}\ }\textbf {\bibinfo {volume} {4}},\
  \bibinfo {pages} {031039} (\bibinfo {year} {2014})}\BibitemShut {NoStop}%
\bibitem [{\citenamefont {Budich}\ \emph {et~al.}(2015)\citenamefont {Budich},
  \citenamefont {Zoller},\ and\ \citenamefont {Diehl}}]{Budich2015}%
  \BibitemOpen
  \bibfield  {author} {\bibinfo {author} {\bibfnamefont {J.~C.}\ \bibnamefont
  {Budich}}, \bibinfo {author} {\bibfnamefont {P.}~\bibnamefont {Zoller}},\
  and\ \bibinfo {author} {\bibfnamefont {S.}~\bibnamefont {Diehl}},\ }\bibfield
   {title} {\bibinfo {title} {Dissipative preparation of {C}hern insulators},\
  }\href {https://www.doi.org/10.1103/physreva.91.042117} {\bibfield  {journal}
  {\bibinfo  {journal} {Physical Review A}\ }\textbf {\bibinfo {volume} {91}},\
  \bibinfo {pages} {042117} (\bibinfo {year} {2015})}\BibitemShut {NoStop}%
\bibitem [{\citenamefont {Iemini}\ \emph {et~al.}(2016)\citenamefont {Iemini},
  \citenamefont {Rossini}, \citenamefont {Fazio}, \citenamefont {Diehl},\ and\
  \citenamefont {Mazza}}]{Iemini2016}%
  \BibitemOpen
  \bibfield  {author} {\bibinfo {author} {\bibfnamefont {F.}~\bibnamefont
  {Iemini}}, \bibinfo {author} {\bibfnamefont {D.}~\bibnamefont {Rossini}},
  \bibinfo {author} {\bibfnamefont {R.}~\bibnamefont {Fazio}}, \bibinfo
  {author} {\bibfnamefont {S.}~\bibnamefont {Diehl}},\ and\ \bibinfo {author}
  {\bibfnamefont {L.}~\bibnamefont {Mazza}},\ }\bibfield  {title} {\bibinfo
  {title} {Dissipative topological superconductors in number-conserving
  systems},\ }\href {https://www.doi.org/10.1103/physrevb.93.115113} {\bibfield
   {journal} {\bibinfo  {journal} {Physical Review B}\ }\textbf {\bibinfo
  {volume} {93}},\ \bibinfo {pages} {115113} (\bibinfo {year}
  {2016})}\BibitemShut {NoStop}%
\bibitem [{\citenamefont {Gong}\ \emph {et~al.}(2017)\citenamefont {Gong},
  \citenamefont {Higashikawa},\ and\ \citenamefont {Ueda}}]{Gong2017}%
  \BibitemOpen
  \bibfield  {author} {\bibinfo {author} {\bibfnamefont {Z.}~\bibnamefont
  {Gong}}, \bibinfo {author} {\bibfnamefont {S.}~\bibnamefont {Higashikawa}},\
  and\ \bibinfo {author} {\bibfnamefont {M.}~\bibnamefont {Ueda}},\ }\bibfield
  {title} {\bibinfo {title} {Zeno {H}all effect},\ }\href
  {https://www.doi.org/10.1103/physrevlett.118.200401} {\bibfield  {journal}
  {\bibinfo  {journal} {Physical Review Letters}\ }\textbf {\bibinfo {volume}
  {118}},\ \bibinfo {pages} {200401} (\bibinfo {year} {2017})}\BibitemShut
  {NoStop}%
\bibitem [{\citenamefont {Goldstein}(2019)}]{Goldstein2019}%
  \BibitemOpen
  \bibfield  {author} {\bibinfo {author} {\bibfnamefont {M.}~\bibnamefont
  {Goldstein}},\ }\bibfield  {title} {\bibinfo {title} {Dissipation-induced
  topological insulators: A no-go theorem and a recipe},\ }\href
  {https://www.doi.org/10.21468/scipostphys.7.5.067} {\bibfield  {journal}
  {\bibinfo  {journal} {{SciPost} Physics}\ }\textbf {\bibinfo {volume} {7}},\
  \bibinfo {pages} {67} (\bibinfo {year} {2019})}\BibitemShut {NoStop}%
\bibitem [{\citenamefont {Shavit}\ and\ \citenamefont
  {Goldstein}(2020)}]{Shavit2020}%
  \BibitemOpen
  \bibfield  {author} {\bibinfo {author} {\bibfnamefont {G.}~\bibnamefont
  {Shavit}}\ and\ \bibinfo {author} {\bibfnamefont {M.}~\bibnamefont
  {Goldstein}},\ }\bibfield  {title} {\bibinfo {title} {Topology by
  dissipation: Transport properties},\ }\href
  {https://www.doi.org/10.1103/physrevb.101.125412} {\bibfield  {journal}
  {\bibinfo  {journal} {Physical Review B}\ }\textbf {\bibinfo {volume}
  {101}},\ \bibinfo {pages} {125412} (\bibinfo {year} {2020})}\BibitemShut
  {NoStop}%
\bibitem [{\citenamefont {Tonielli}\ \emph {et~al.}(2020)\citenamefont
  {Tonielli}, \citenamefont {Budich}, \citenamefont {Altland},\ and\
  \citenamefont {Diehl}}]{Tonielli2020}%
  \BibitemOpen
  \bibfield  {author} {\bibinfo {author} {\bibfnamefont {F.}~\bibnamefont
  {Tonielli}}, \bibinfo {author} {\bibfnamefont {J.~C.}\ \bibnamefont
  {Budich}}, \bibinfo {author} {\bibfnamefont {A.}~\bibnamefont {Altland}},\
  and\ \bibinfo {author} {\bibfnamefont {S.}~\bibnamefont {Diehl}},\ }\bibfield
   {title} {\bibinfo {title} {Topological field theory far from equilibrium},\
  }\href {https://doi.org/10.1103/PhysRevLett.124.240404} {\bibfield  {journal}
  {\bibinfo  {journal} {Phys. Rev. Lett.}\ }\textbf {\bibinfo {volume} {124}},\
  \bibinfo {pages} {240404} (\bibinfo {year} {2020})}\BibitemShut {NoStop}%
\bibitem [{\citenamefont {Yoshida}\ \emph {et~al.}(2020)\citenamefont
  {Yoshida}, \citenamefont {Kudo}, \citenamefont {Katsura},\ and\ \citenamefont
  {Hatsugai}}]{Yoshida2020}%
  \BibitemOpen
  \bibfield  {author} {\bibinfo {author} {\bibfnamefont {T.}~\bibnamefont
  {Yoshida}}, \bibinfo {author} {\bibfnamefont {K.}~\bibnamefont {Kudo}},
  \bibinfo {author} {\bibfnamefont {H.}~\bibnamefont {Katsura}},\ and\ \bibinfo
  {author} {\bibfnamefont {Y.}~\bibnamefont {Hatsugai}},\ }\bibfield  {title}
  {\bibinfo {title} {Fate of fractional quantum {H}all states in open quantum
  systems: Characterization of correlated topological states for the full
  {L}iouvillian},\ }\href {https://doi.org/10.1103/PhysRevResearch.2.033428}
  {\bibfield  {journal} {\bibinfo  {journal} {Phys. Rev. Research}\ }\textbf
  {\bibinfo {volume} {2}},\ \bibinfo {pages} {033428} (\bibinfo {year}
  {2020})}\BibitemShut {NoStop}%
\bibitem [{\citenamefont {Bandyopadhyay}\ and\ \citenamefont
  {Dutta}(2020)}]{Bandyopadhyay2020}%
  \BibitemOpen
  \bibfield  {author} {\bibinfo {author} {\bibfnamefont {S.}~\bibnamefont
  {Bandyopadhyay}}\ and\ \bibinfo {author} {\bibfnamefont {A.}~\bibnamefont
  {Dutta}},\ }\bibfield  {title} {\bibinfo {title} {Dissipative preparation of
  many-body {F}loquet {C}hern insulators},\ }\Eprint
  {https://arxiv.org/abs/2005.09972} {arXiv:2005.09972 [cond-mat.stat-mech]}
  (\bibinfo {year} {2020})\BibitemShut {NoStop}%
\bibitem [{\citenamefont {Altland}\ \emph {et~al.}(2020)\citenamefont
  {Altland}, \citenamefont {Fleischhauer},\ and\ \citenamefont
  {Diehl}}]{Altland2020}%
  \BibitemOpen
  \bibfield  {author} {\bibinfo {author} {\bibfnamefont {A.}~\bibnamefont
  {Altland}}, \bibinfo {author} {\bibfnamefont {M.}~\bibnamefont
  {Fleischhauer}},\ and\ \bibinfo {author} {\bibfnamefont {S.}~\bibnamefont
  {Diehl}},\ }\bibfield  {title} {\bibinfo {title} {Symmetry classes of open
  fermionic quantum matter},\ }\Eprint {https://arxiv.org/abs/2007.10448}
  {arXiv:2007.10448 [cond-mat.str-el]}  (\bibinfo {year} {2020})\BibitemShut
  {NoStop}%
\bibitem [{\citenamefont {Rivas}\ \emph {et~al.}(2013)\citenamefont {Rivas},
  \citenamefont {Viyuela},\ and\ \citenamefont {Martin-Delgado}}]{Rivas2013}%
  \BibitemOpen
  \bibfield  {author} {\bibinfo {author} {\bibfnamefont {A.}~\bibnamefont
  {Rivas}}, \bibinfo {author} {\bibfnamefont {O.}~\bibnamefont {Viyuela}},\
  and\ \bibinfo {author} {\bibfnamefont {M.~A.}\ \bibnamefont
  {Martin-Delgado}},\ }\bibfield  {title} {\bibinfo {title} {Density-matrix
  {C}hern insulators: Finite-temperature generalization of topological
  insulators},\ }\href {https://doi.org/10.1103/PhysRevB.88.155141} {\bibfield
  {journal} {\bibinfo  {journal} {Phys. Rev. B}\ }\textbf {\bibinfo {volume}
  {88}},\ \bibinfo {pages} {155141} (\bibinfo {year} {2013})}\BibitemShut
  {NoStop}%
\bibitem [{\citenamefont {Huang}\ and\ \citenamefont
  {Arovas}(2014)}]{Huang2014}%
  \BibitemOpen
  \bibfield  {author} {\bibinfo {author} {\bibfnamefont {Z.}~\bibnamefont
  {Huang}}\ and\ \bibinfo {author} {\bibfnamefont {D.~P.}\ \bibnamefont
  {Arovas}},\ }\bibfield  {title} {\bibinfo {title} {Topological indices for
  open and thermal systems via {U}hlmann's phase},\ }\href
  {https://doi.org/10.1103/PhysRevLett.113.076407} {\bibfield  {journal}
  {\bibinfo  {journal} {Phys. Rev. Lett.}\ }\textbf {\bibinfo {volume} {113}},\
  \bibinfo {pages} {076407} (\bibinfo {year} {2014})}\BibitemShut {NoStop}%
\bibitem [{\citenamefont {Viyuela}\ \emph {et~al.}(2014)\citenamefont
  {Viyuela}, \citenamefont {Rivas},\ and\ \citenamefont
  {Martin-Delgado}}]{Viyuela2014}%
  \BibitemOpen
  \bibfield  {author} {\bibinfo {author} {\bibfnamefont {O.}~\bibnamefont
  {Viyuela}}, \bibinfo {author} {\bibfnamefont {A.}~\bibnamefont {Rivas}},\
  and\ \bibinfo {author} {\bibfnamefont {M.~A.}\ \bibnamefont
  {Martin-Delgado}},\ }\bibfield  {title} {\bibinfo {title} {Two-dimensional
  density-matrix topological fermionic phases: Topological {U}hlmann numbers},\
  }\href {https://doi.org/10.1103/PhysRevLett.113.076408} {\bibfield  {journal}
  {\bibinfo  {journal} {Phys. Rev. Lett.}\ }\textbf {\bibinfo {volume} {113}},\
  \bibinfo {pages} {076408} (\bibinfo {year} {2014})}\BibitemShut {NoStop}%
\bibitem [{\citenamefont {van Nieuwenburg}\ and\ \citenamefont
  {Huber}(2014)}]{Nieuwenburg2014}%
  \BibitemOpen
  \bibfield  {author} {\bibinfo {author} {\bibfnamefont {E.~P.~L.}\
  \bibnamefont {van Nieuwenburg}}\ and\ \bibinfo {author} {\bibfnamefont
  {S.~D.}\ \bibnamefont {Huber}},\ }\bibfield  {title} {\bibinfo {title}
  {Classification of mixed-state topology in one dimension},\ }\href
  {https://doi.org/10.1103/PhysRevB.90.075141} {\bibfield  {journal} {\bibinfo
  {journal} {Phys. Rev. B}\ }\textbf {\bibinfo {volume} {90}},\ \bibinfo
  {pages} {075141} (\bibinfo {year} {2014})}\BibitemShut {NoStop}%
\bibitem [{\citenamefont {Budich}\ and\ \citenamefont
  {Diehl}(2015)}]{Budich2015b}%
  \BibitemOpen
  \bibfield  {author} {\bibinfo {author} {\bibfnamefont {J.~C.}\ \bibnamefont
  {Budich}}\ and\ \bibinfo {author} {\bibfnamefont {S.}~\bibnamefont {Diehl}},\
  }\bibfield  {title} {\bibinfo {title} {Topology of density matrices},\ }\href
  {https://doi.org/10.1103/PhysRevB.91.165140} {\bibfield  {journal} {\bibinfo
  {journal} {Phys. Rev. B}\ }\textbf {\bibinfo {volume} {91}},\ \bibinfo
  {pages} {165140} (\bibinfo {year} {2015})}\BibitemShut {NoStop}%
\bibitem [{\citenamefont {Grusdt}(2017)}]{Grusdt2017}%
  \BibitemOpen
  \bibfield  {author} {\bibinfo {author} {\bibfnamefont {F.}~\bibnamefont
  {Grusdt}},\ }\bibfield  {title} {\bibinfo {title} {Topological order of mixed
  states in correlated quantum many-body systems},\ }\href
  {https://doi.org/10.1103/PhysRevB.95.075106} {\bibfield  {journal} {\bibinfo
  {journal} {Phys. Rev. B}\ }\textbf {\bibinfo {volume} {95}},\ \bibinfo
  {pages} {075106} (\bibinfo {year} {2017})}\BibitemShut {NoStop}%
\bibitem [{\citenamefont {Bardyn}(2017)}]{Bardyn2017}%
  \BibitemOpen
  \bibfield  {author} {\bibinfo {author} {\bibfnamefont {C.-E.}\ \bibnamefont
  {Bardyn}},\ }\bibfield  {title} {\bibinfo {title} {A recipe for topological
  observables of density matrices},\ }\Eprint
  {https://arxiv.org/abs/1711.09735} {arXiv:1711.09735 [cond-mat.quant-gas]}
  (\bibinfo {year} {2017})\BibitemShut {NoStop}%
\bibitem [{\citenamefont {Bardyn}\ \emph {et~al.}(2018)\citenamefont {Bardyn},
  \citenamefont {Wawer}, \citenamefont {Altland}, \citenamefont
  {Fleischhauer},\ and\ \citenamefont {Diehl}}]{Bardyn2018}%
  \BibitemOpen
  \bibfield  {author} {\bibinfo {author} {\bibfnamefont {C.-E.}\ \bibnamefont
  {Bardyn}}, \bibinfo {author} {\bibfnamefont {L.}~\bibnamefont {Wawer}},
  \bibinfo {author} {\bibfnamefont {A.}~\bibnamefont {Altland}}, \bibinfo
  {author} {\bibfnamefont {M.}~\bibnamefont {Fleischhauer}},\ and\ \bibinfo
  {author} {\bibfnamefont {S.}~\bibnamefont {Diehl}},\ }\bibfield  {title}
  {\bibinfo {title} {Probing the topology of density matrices},\ }\href
  {https://doi.org/10.1103/PhysRevX.8.011035} {\bibfield  {journal} {\bibinfo
  {journal} {Phys. Rev. X}\ }\textbf {\bibinfo {volume} {8}},\ \bibinfo {pages}
  {011035} (\bibinfo {year} {2018})}\BibitemShut {NoStop}%
\bibitem [{\citenamefont {Zhang}\ and\ \citenamefont {Gong}(2018)}]{Zhang2018}%
  \BibitemOpen
  \bibfield  {author} {\bibinfo {author} {\bibfnamefont {D.-J.}\ \bibnamefont
  {Zhang}}\ and\ \bibinfo {author} {\bibfnamefont {J.}~\bibnamefont {Gong}},\
  }\bibfield  {title} {\bibinfo {title} {Topological characterization of
  one-dimensional open fermionic systems},\ }\href
  {https://doi.org/10.1103/PhysRevA.98.052101} {\bibfield  {journal} {\bibinfo
  {journal} {Phys. Rev. A}\ }\textbf {\bibinfo {volume} {98}},\ \bibinfo
  {pages} {052101} (\bibinfo {year} {2018})}\BibitemShut {NoStop}%
\bibitem [{\citenamefont {Coser}\ and\ \citenamefont
  {P{\'{e}}rez-Garc{\'{i}}a}(2019)}]{Coser2019}%
  \BibitemOpen
  \bibfield  {author} {\bibinfo {author} {\bibfnamefont {A.}~\bibnamefont
  {Coser}}\ and\ \bibinfo {author} {\bibfnamefont {D.}~\bibnamefont
  {P{\'{e}}rez-Garc{\'{i}}a}},\ }\bibfield  {title} {\bibinfo {title}
  {Classification of phases for mixed states via fast dissipative evolution},\
  }\href {https://doi.org/10.22331/q-2019-08-12-174} {\bibfield  {journal}
  {\bibinfo  {journal} {{Quantum}}\ }\textbf {\bibinfo {volume} {3}},\ \bibinfo
  {pages} {174} (\bibinfo {year} {2019})}\BibitemShut {NoStop}%
\bibitem [{\citenamefont {Lieu}\ \emph {et~al.}(2020)\citenamefont {Lieu},
  \citenamefont {McGinley},\ and\ \citenamefont {Cooper}}]{Lieu2020}%
  \BibitemOpen
  \bibfield  {author} {\bibinfo {author} {\bibfnamefont {S.}~\bibnamefont
  {Lieu}}, \bibinfo {author} {\bibfnamefont {M.}~\bibnamefont {McGinley}},\
  and\ \bibinfo {author} {\bibfnamefont {N.~R.}\ \bibnamefont {Cooper}},\
  }\bibfield  {title} {\bibinfo {title} {Tenfold way for quadratic
  {L}indbladians},\ }\href {https://doi.org/10.1103/PhysRevLett.124.040401}
  {\bibfield  {journal} {\bibinfo  {journal} {Phys. Rev. Lett.}\ }\textbf
  {\bibinfo {volume} {124}},\ \bibinfo {pages} {040401} (\bibinfo {year}
  {2020})}\BibitemShut {NoStop}%
\bibitem [{\citenamefont {v.~Klitzing}\ \emph {et~al.}(1980)\citenamefont
  {v.~Klitzing}, \citenamefont {Dorda},\ and\ \citenamefont
  {Pepper}}]{Klitzing1980}%
  \BibitemOpen
  \bibfield  {author} {\bibinfo {author} {\bibfnamefont {K.}~\bibnamefont
  {v.~Klitzing}}, \bibinfo {author} {\bibfnamefont {G.}~\bibnamefont {Dorda}},\
  and\ \bibinfo {author} {\bibfnamefont {M.}~\bibnamefont {Pepper}},\
  }\bibfield  {title} {\bibinfo {title} {New method for high-accuracy
  determination of the fine-structure constant based on quantized {H}all
  resistance},\ }\href {https://doi.org/10.1103/PhysRevLett.45.494} {\bibfield
  {journal} {\bibinfo  {journal} {Phys. Rev. Lett.}\ }\textbf {\bibinfo
  {volume} {45}},\ \bibinfo {pages} {494} (\bibinfo {year} {1980})}\BibitemShut
  {NoStop}%
\bibitem [{\citenamefont {Huckestein}(1995)}]{Huckestein1995}%
  \BibitemOpen
  \bibfield  {author} {\bibinfo {author} {\bibfnamefont {B.}~\bibnamefont
  {Huckestein}},\ }\bibfield  {title} {\bibinfo {title} {Scaling theory of the
  integer quantum {H}all effect},\ }\href {10.1103/revmodphys.67.357}
  {\bibfield  {journal} {\bibinfo  {journal} {Reviews of Modern Physics}\
  }\textbf {\bibinfo {volume} {67}},\ \bibinfo {pages} {357} (\bibinfo {year}
  {1995})}\BibitemShut {NoStop}%
\bibitem [{\citenamefont {Evers}\ and\ \citenamefont
  {Mirlin}(2008)}]{Evers2008}%
  \BibitemOpen
  \bibfield  {author} {\bibinfo {author} {\bibfnamefont {F.}~\bibnamefont
  {Evers}}\ and\ \bibinfo {author} {\bibfnamefont {A.~D.}\ \bibnamefont
  {Mirlin}},\ }\bibfield  {title} {\bibinfo {title} {Anderson transitions},\
  }\href {https://www.doi.org/10.1103/revmodphys.80.1355} {\bibfield  {journal}
  {\bibinfo  {journal} {Reviews of Modern Physics}\ }\textbf {\bibinfo {volume}
  {80}},\ \bibinfo {pages} {1355} (\bibinfo {year} {2008})}\BibitemShut
  {NoStop}%
\bibitem [{\citenamefont {Slevin}\ and\ \citenamefont
  {Ohtsuki}(2009)}]{Slevin2009}%
  \BibitemOpen
  \bibfield  {author} {\bibinfo {author} {\bibfnamefont {K.}~\bibnamefont
  {Slevin}}\ and\ \bibinfo {author} {\bibfnamefont {T.}~\bibnamefont
  {Ohtsuki}},\ }\bibfield  {title} {\bibinfo {title} {Critical exponent for the
  quantum hall transition},\ }\href
  {https://doi.org/10.1103/PhysRevB.80.041304} {\bibfield  {journal} {\bibinfo
  {journal} {Phys. Rev. B}\ }\textbf {\bibinfo {volume} {80}},\ \bibinfo
  {pages} {041304} (\bibinfo {year} {2009})}\BibitemShut {NoStop}%
\bibitem [{\citenamefont {Obuse}\ \emph {et~al.}(2010)\citenamefont {Obuse},
  \citenamefont {Subramaniam}, \citenamefont {Furusaki}, \citenamefont
  {Gruzberg},\ and\ \citenamefont {Ludwig}}]{Obuse2010}%
  \BibitemOpen
  \bibfield  {author} {\bibinfo {author} {\bibfnamefont {H.}~\bibnamefont
  {Obuse}}, \bibinfo {author} {\bibfnamefont {A.~R.}\ \bibnamefont
  {Subramaniam}}, \bibinfo {author} {\bibfnamefont {A.}~\bibnamefont
  {Furusaki}}, \bibinfo {author} {\bibfnamefont {I.~A.}\ \bibnamefont
  {Gruzberg}},\ and\ \bibinfo {author} {\bibfnamefont {A.~W.~W.}\ \bibnamefont
  {Ludwig}},\ }\bibfield  {title} {\bibinfo {title} {Conformal invariance,
  multifractality, and finite-size scaling at anderson localization transitions
  in two dimensions},\ }\href {https://www.doi.org/10.1103/physrevb.82.035309}
  {\bibfield  {journal} {\bibinfo  {journal} {Physical Review B}\ }\textbf
  {\bibinfo {volume} {82}},\ \bibinfo {pages} {035309} (\bibinfo {year}
  {2010})}\BibitemShut {NoStop}%
\bibitem [{\citenamefont {Amado}\ \emph {et~al.}(2011)\citenamefont {Amado},
  \citenamefont {Malyshev}, \citenamefont {Sedrakyan},\ and\ \citenamefont
  {Dom{\'{\i}}nguez-Adame}}]{Amado2011}%
  \BibitemOpen
  \bibfield  {author} {\bibinfo {author} {\bibfnamefont {M.}~\bibnamefont
  {Amado}}, \bibinfo {author} {\bibfnamefont {A.~V.}\ \bibnamefont {Malyshev}},
  \bibinfo {author} {\bibfnamefont {A.}~\bibnamefont {Sedrakyan}},\ and\
  \bibinfo {author} {\bibfnamefont {F.}~\bibnamefont
  {Dom{\'{\i}}nguez-Adame}},\ }\bibfield  {title} {\bibinfo {title} {Numerical
  study of the localization length critical index in a network model of
  plateau-plateau transitions in the quantum {H}all effect},\ }\href
  {https://www.doi.org/10.1103/physrevlett.107.066402} {\bibfield  {journal}
  {\bibinfo  {journal} {Physical Review Letters}\ }\textbf {\bibinfo {volume}
  {107}},\ \bibinfo {pages} {066402} (\bibinfo {year} {2011})}\BibitemShut
  {NoStop}%
\bibitem [{\citenamefont {Fulga}\ \emph {et~al.}(2011)\citenamefont {Fulga},
  \citenamefont {Hassler}, \citenamefont {Akhmerov},\ and\ \citenamefont
  {Beenakker}}]{Fulga2011}%
  \BibitemOpen
  \bibfield  {author} {\bibinfo {author} {\bibfnamefont {I.~C.}\ \bibnamefont
  {Fulga}}, \bibinfo {author} {\bibfnamefont {F.}~\bibnamefont {Hassler}},
  \bibinfo {author} {\bibfnamefont {A.~R.}\ \bibnamefont {Akhmerov}},\ and\
  \bibinfo {author} {\bibfnamefont {C.~W.~J.}\ \bibnamefont {Beenakker}},\
  }\bibfield  {title} {\bibinfo {title} {Topological quantum number and
  critical exponent from conductance fluctuations at the quantum {H}all plateau
  transition},\ }\href {https://www.doi.org/10.1103/physrevb.84.245447}
  {\bibfield  {journal} {\bibinfo  {journal} {Physical Review B}\ }\textbf
  {\bibinfo {volume} {84}},\ \bibinfo {pages} {245447} (\bibinfo {year}
  {2011})}\BibitemShut {NoStop}%
\bibitem [{\citenamefont {Slevin}\ and\ \citenamefont
  {Ohtsuki}(2012)}]{Slevin2012}%
  \BibitemOpen
  \bibfield  {author} {\bibinfo {author} {\bibfnamefont {K.}~\bibnamefont
  {Slevin}}\ and\ \bibinfo {author} {\bibfnamefont {T.}~\bibnamefont
  {Ohtsuki}},\ }\bibfield  {title} {\bibinfo {title} {Finite size scaling of
  the {C}halker-{C}oddington model},\ }\href
  {https://www.doi.org/10.1142/s2010194512006162} {\bibfield  {journal}
  {\bibinfo  {journal} {International Journal of Modern Physics: Conference
  Series}\ }\textbf {\bibinfo {volume} {11}},\ \bibinfo {pages} {60} (\bibinfo
  {year} {2012})}\BibitemShut {NoStop}%
\bibitem [{\citenamefont {Obuse}\ \emph {et~al.}(2012)\citenamefont {Obuse},
  \citenamefont {Gruzberg},\ and\ \citenamefont {Evers}}]{Obuse2012}%
  \BibitemOpen
  \bibfield  {author} {\bibinfo {author} {\bibfnamefont {H.}~\bibnamefont
  {Obuse}}, \bibinfo {author} {\bibfnamefont {I.~A.}\ \bibnamefont
  {Gruzberg}},\ and\ \bibinfo {author} {\bibfnamefont {F.}~\bibnamefont
  {Evers}},\ }\bibfield  {title} {\bibinfo {title} {Finite-size effects and
  irrelevant corrections to scaling near the integer quantum {H}all
  transition},\ }\href {https://www.doi.org/10.1103/physrevlett.109.206804}
  {\bibfield  {journal} {\bibinfo  {journal} {Physical Review Letters}\
  }\textbf {\bibinfo {volume} {109}},\ \bibinfo {pages} {206804} (\bibinfo
  {year} {2012})}\BibitemShut {NoStop}%
\bibitem [{\citenamefont {Nuding}\ \emph {et~al.}(2015)\citenamefont {Nuding},
  \citenamefont {Klumper},\ and\ \citenamefont {Sedrakyan}}]{Nuding2015}%
  \BibitemOpen
  \bibfield  {author} {\bibinfo {author} {\bibfnamefont {W.}~\bibnamefont
  {Nuding}}, \bibinfo {author} {\bibfnamefont {A.}~\bibnamefont {Klumper}},\
  and\ \bibinfo {author} {\bibfnamefont {A.}~\bibnamefont {Sedrakyan}},\
  }\bibfield  {title} {\bibinfo {title} {Localization length index and
  subleading corrections in a {C}halker-{C}oddington model: A numerical
  study},\ }\href {https://www.doi.org/10.1103/physrevb.91.115107} {\bibfield
  {journal} {\bibinfo  {journal} {Physical Review B}\ }\textbf {\bibinfo
  {volume} {91}},\ \bibinfo {pages} {115107} (\bibinfo {year}
  {2015})}\BibitemShut {NoStop}%
\bibitem [{\citenamefont {Gruzberg}\ \emph {et~al.}(2017)\citenamefont
  {Gruzberg}, \citenamefont {Klumper}, \citenamefont {Nuding},\ and\
  \citenamefont {Sedrakyan}}]{Gruzberg2017}%
  \BibitemOpen
  \bibfield  {author} {\bibinfo {author} {\bibfnamefont {I.~A.}\ \bibnamefont
  {Gruzberg}}, \bibinfo {author} {\bibfnamefont {A.}~\bibnamefont {Klumper}},
  \bibinfo {author} {\bibfnamefont {W.}~\bibnamefont {Nuding}},\ and\ \bibinfo
  {author} {\bibfnamefont {A.}~\bibnamefont {Sedrakyan}},\ }\bibfield  {title}
  {\bibinfo {title} {Geometrically disordered network models, quenched quantum
  gravity, and critical behavior at quantum {H}all plateau transitions},\
  }\href {https://www.doi.org/10.1103/physrevb.95.125414} {\bibfield  {journal}
  {\bibinfo  {journal} {Physical Review B}\ }\textbf {\bibinfo {volume} {95}},\
  \bibinfo {pages} {125414} (\bibinfo {year} {2017})}\BibitemShut {NoStop}%
\bibitem [{\citenamefont {Ippoliti}\ \emph {et~al.}(2018)\citenamefont
  {Ippoliti}, \citenamefont {Geraedts},\ and\ \citenamefont
  {Bhatt}}]{Ippoliti2018}%
  \BibitemOpen
  \bibfield  {author} {\bibinfo {author} {\bibfnamefont {M.}~\bibnamefont
  {Ippoliti}}, \bibinfo {author} {\bibfnamefont {S.~D.}\ \bibnamefont
  {Geraedts}},\ and\ \bibinfo {author} {\bibfnamefont {R.~N.}\ \bibnamefont
  {Bhatt}},\ }\bibfield  {title} {\bibinfo {title} {Integer quantum {H}all
  transition in a fraction of a landau level},\ }\href
  {https://www.doi.org/10.1103/physrevb.97.014205} {\bibfield  {journal}
  {\bibinfo  {journal} {Physical Review B}\ }\textbf {\bibinfo {volume} {97}},\
  \bibinfo {pages} {014205} (\bibinfo {year} {2018})}\BibitemShut {NoStop}%
\bibitem [{\citenamefont {Puschmann}\ \emph {et~al.}(2019)\citenamefont
  {Puschmann}, \citenamefont {Cain}, \citenamefont {Schreiber},\ and\
  \citenamefont {Vojta}}]{Puschmann2019}%
  \BibitemOpen
  \bibfield  {author} {\bibinfo {author} {\bibfnamefont {M.}~\bibnamefont
  {Puschmann}}, \bibinfo {author} {\bibfnamefont {P.}~\bibnamefont {Cain}},
  \bibinfo {author} {\bibfnamefont {M.}~\bibnamefont {Schreiber}},\ and\
  \bibinfo {author} {\bibfnamefont {T.}~\bibnamefont {Vojta}},\ }\bibfield
  {title} {\bibinfo {title} {Integer quantum {H}all transition on a
  tight-binding lattice},\ }\href
  {https://www.doi.org/10.1103/physrevb.99.121301} {\bibfield  {journal}
  {\bibinfo  {journal} {Physical Review B}\ }\textbf {\bibinfo {volume} {99}},\
  \bibinfo {pages} {121301} (\bibinfo {year} {2019})}\BibitemShut {NoStop}%
\bibitem [{\citenamefont {Zhu}\ \emph {et~al.}(2019)\citenamefont {Zhu},
  \citenamefont {Wu}, \citenamefont {Bhatt},\ and\ \citenamefont
  {Wan}}]{Zhu2019}%
  \BibitemOpen
  \bibfield  {author} {\bibinfo {author} {\bibfnamefont {Q.}~\bibnamefont
  {Zhu}}, \bibinfo {author} {\bibfnamefont {P.}~\bibnamefont {Wu}}, \bibinfo
  {author} {\bibfnamefont {R.~N.}\ \bibnamefont {Bhatt}},\ and\ \bibinfo
  {author} {\bibfnamefont {X.}~\bibnamefont {Wan}},\ }\bibfield  {title}
  {\bibinfo {title} {Localization-length exponent in two models of quantum
  {H}all plateau transitions},\ }\href
  {https://www.doi.org/10.1103/physrevb.99.024205} {\bibfield  {journal}
  {\bibinfo  {journal} {Physical Review B}\ }\textbf {\bibinfo {volume} {99}},\
  \bibinfo {pages} {024205} (\bibinfo {year} {2019})}\BibitemShut {NoStop}%
\bibitem [{\citenamefont {Sbierski}\ \emph {et~al.}(2021)\citenamefont
  {Sbierski}, \citenamefont {Dresselhaus}, \citenamefont {Moore},\ and\
  \citenamefont {Gruzberg}}]{Sbierski2021}%
  \BibitemOpen
  \bibfield  {author} {\bibinfo {author} {\bibfnamefont {B.}~\bibnamefont
  {Sbierski}}, \bibinfo {author} {\bibfnamefont {E.~J.}\ \bibnamefont
  {Dresselhaus}}, \bibinfo {author} {\bibfnamefont {J.~E.}\ \bibnamefont
  {Moore}},\ and\ \bibinfo {author} {\bibfnamefont {I.~A.}\ \bibnamefont
  {Gruzberg}},\ }\bibfield  {title} {\bibinfo {title} {Criticality of
  two-dimensional disordered dirac fermions in the unitary class and
  universality of the integer quantum hall transition},\ }\href
  {https://doi.org/10.1103/PhysRevLett.126.076801} {\bibfield  {journal}
  {\bibinfo  {journal} {Phys. Rev. Lett.}\ }\textbf {\bibinfo {volume} {126}},\
  \bibinfo {pages} {076801} (\bibinfo {year} {2021})}\BibitemShut {NoStop}%
\bibitem [{\citenamefont {Li}\ \emph {et~al.}(2005)\citenamefont {Li},
  \citenamefont {Cs\'athy}, \citenamefont {Tsui}, \citenamefont {Pfeiffer},\
  and\ \citenamefont {West}}]{Li2005}%
  \BibitemOpen
  \bibfield  {author} {\bibinfo {author} {\bibfnamefont {W.}~\bibnamefont
  {Li}}, \bibinfo {author} {\bibfnamefont {G.~A.}\ \bibnamefont {Cs\'athy}},
  \bibinfo {author} {\bibfnamefont {D.~C.}\ \bibnamefont {Tsui}}, \bibinfo
  {author} {\bibfnamefont {L.~N.}\ \bibnamefont {Pfeiffer}},\ and\ \bibinfo
  {author} {\bibfnamefont {K.~W.}\ \bibnamefont {West}},\ }\bibfield  {title}
  {\bibinfo {title} {Scaling and universality of integer quantum {H}all
  plateau-to-plateau transitions},\ }\href
  {https://doi.org/10.1103/PhysRevLett.94.206807} {\bibfield  {journal}
  {\bibinfo  {journal} {Phys. Rev. Lett.}\ }\textbf {\bibinfo {volume} {94}},\
  \bibinfo {pages} {206807} (\bibinfo {year} {2005})}\BibitemShut {NoStop}%
\bibitem [{\citenamefont {Li}\ \emph {et~al.}(2009)\citenamefont {Li},
  \citenamefont {Vicente}, \citenamefont {Xia}, \citenamefont {Pan},
  \citenamefont {Tsui}, \citenamefont {Pfeiffer},\ and\ \citenamefont
  {West}}]{Li2009}%
  \BibitemOpen
  \bibfield  {author} {\bibinfo {author} {\bibfnamefont {W.}~\bibnamefont
  {Li}}, \bibinfo {author} {\bibfnamefont {C.~L.}\ \bibnamefont {Vicente}},
  \bibinfo {author} {\bibfnamefont {J.~S.}\ \bibnamefont {Xia}}, \bibinfo
  {author} {\bibfnamefont {W.}~\bibnamefont {Pan}}, \bibinfo {author}
  {\bibfnamefont {D.~C.}\ \bibnamefont {Tsui}}, \bibinfo {author}
  {\bibfnamefont {L.~N.}\ \bibnamefont {Pfeiffer}},\ and\ \bibinfo {author}
  {\bibfnamefont {K.~W.}\ \bibnamefont {West}},\ }\bibfield  {title} {\bibinfo
  {title} {Scaling in plateau-to-plateau transition: A direct connection of
  quantum {H}all systems with the anderson localization model},\ }\href
  {https://doi.org/10.1103/PhysRevLett.102.216801} {\bibfield  {journal}
  {\bibinfo  {journal} {Phys. Rev. Lett.}\ }\textbf {\bibinfo {volume} {102}},\
  \bibinfo {pages} {216801} (\bibinfo {year} {2009})}\BibitemShut {NoStop}%
\bibitem [{\citenamefont {Giesbers}\ \emph {et~al.}(2009)\citenamefont
  {Giesbers}, \citenamefont {Zeitler}, \citenamefont {Ponomarenko},
  \citenamefont {Yang}, \citenamefont {Novoselov}, \citenamefont {Geim},\ and\
  \citenamefont {Maan}}]{Giesbers2009}%
  \BibitemOpen
  \bibfield  {author} {\bibinfo {author} {\bibfnamefont {A.~J.~M.}\
  \bibnamefont {Giesbers}}, \bibinfo {author} {\bibfnamefont {U.}~\bibnamefont
  {Zeitler}}, \bibinfo {author} {\bibfnamefont {L.~A.}\ \bibnamefont
  {Ponomarenko}}, \bibinfo {author} {\bibfnamefont {R.}~\bibnamefont {Yang}},
  \bibinfo {author} {\bibfnamefont {K.~S.}\ \bibnamefont {Novoselov}}, \bibinfo
  {author} {\bibfnamefont {A.~K.}\ \bibnamefont {Geim}},\ and\ \bibinfo
  {author} {\bibfnamefont {J.~C.}\ \bibnamefont {Maan}},\ }\bibfield  {title}
  {\bibinfo {title} {Scaling of the quantum {H}all plateau-plateau transition
  in graphene},\ }\href {https://doi.org/10.1103/PhysRevB.80.241411} {\bibfield
   {journal} {\bibinfo  {journal} {Phys. Rev. B}\ }\textbf {\bibinfo {volume}
  {80}},\ \bibinfo {pages} {241411} (\bibinfo {year} {2009})}\BibitemShut
  {NoStop}%
\bibitem [{\citenamefont {Hauke}\ \emph {et~al.}(2014)\citenamefont {Hauke},
  \citenamefont {Lewenstein},\ and\ \citenamefont {Eckardt}}]{Hauke2014}%
  \BibitemOpen
  \bibfield  {author} {\bibinfo {author} {\bibfnamefont {P.}~\bibnamefont
  {Hauke}}, \bibinfo {author} {\bibfnamefont {M.}~\bibnamefont {Lewenstein}},\
  and\ \bibinfo {author} {\bibfnamefont {A.}~\bibnamefont {Eckardt}},\
  }\bibfield  {title} {\bibinfo {title} {Tomography of band insulators from
  quench dynamics},\ }\href {https://doi.org/10.1103/PhysRevLett.113.045303}
  {\bibfield  {journal} {\bibinfo  {journal} {Phys. Rev. Lett.}\ }\textbf
  {\bibinfo {volume} {113}},\ \bibinfo {pages} {045303} (\bibinfo {year}
  {2014})}\BibitemShut {NoStop}%
\bibitem [{\citenamefont {Fl{\"a}schner}\ \emph {et~al.}(2016)\citenamefont
  {Fl{\"a}schner}, \citenamefont {Rem}, \citenamefont {Tarnowski},
  \citenamefont {Vogel}, \citenamefont {L{\"u}hmann}, \citenamefont
  {Sengstock},\ and\ \citenamefont {Weitenberg}}]{Flaschner2016}%
  \BibitemOpen
  \bibfield  {author} {\bibinfo {author} {\bibfnamefont {N.}~\bibnamefont
  {Fl{\"a}schner}}, \bibinfo {author} {\bibfnamefont {B.~S.}\ \bibnamefont
  {Rem}}, \bibinfo {author} {\bibfnamefont {M.}~\bibnamefont {Tarnowski}},
  \bibinfo {author} {\bibfnamefont {D.}~\bibnamefont {Vogel}}, \bibinfo
  {author} {\bibfnamefont {D.-S.}\ \bibnamefont {L{\"u}hmann}}, \bibinfo
  {author} {\bibfnamefont {K.}~\bibnamefont {Sengstock}},\ and\ \bibinfo
  {author} {\bibfnamefont {C.}~\bibnamefont {Weitenberg}},\ }\bibfield  {title}
  {\bibinfo {title} {Experimental reconstruction of the berry curvature in a
  {F}loquet {B}loch band},\ }\href {https://doi.org/10.1126/science.aad4568}
  {\bibfield  {journal} {\bibinfo  {journal} {Science}\ }\textbf {\bibinfo
  {volume} {352}},\ \bibinfo {pages} {1091} (\bibinfo {year}
  {2016})}\BibitemShut {NoStop}%
\bibitem [{\citenamefont {Tarnowski}\ \emph {et~al.}(2017)\citenamefont
  {Tarnowski}, \citenamefont {Nuske}, \citenamefont {Fl\"aschner},
  \citenamefont {Rem}, \citenamefont {Vogel}, \citenamefont {Freystatzky},
  \citenamefont {Sengstock}, \citenamefont {Mathey},\ and\ \citenamefont
  {Weitenberg}}]{Tarnowski2017}%
  \BibitemOpen
  \bibfield  {author} {\bibinfo {author} {\bibfnamefont {M.}~\bibnamefont
  {Tarnowski}}, \bibinfo {author} {\bibfnamefont {M.}~\bibnamefont {Nuske}},
  \bibinfo {author} {\bibfnamefont {N.}~\bibnamefont {Fl\"aschner}}, \bibinfo
  {author} {\bibfnamefont {B.}~\bibnamefont {Rem}}, \bibinfo {author}
  {\bibfnamefont {D.}~\bibnamefont {Vogel}}, \bibinfo {author} {\bibfnamefont
  {L.}~\bibnamefont {Freystatzky}}, \bibinfo {author} {\bibfnamefont
  {K.}~\bibnamefont {Sengstock}}, \bibinfo {author} {\bibfnamefont
  {L.}~\bibnamefont {Mathey}},\ and\ \bibinfo {author} {\bibfnamefont
  {C.}~\bibnamefont {Weitenberg}},\ }\bibfield  {title} {\bibinfo {title}
  {Observation of topological {B}loch-state defects and their merging
  transition},\ }\href {https://doi.org/10.1103/PhysRevLett.118.240403}
  {\bibfield  {journal} {\bibinfo  {journal} {Phys. Rev. Lett.}\ }\textbf
  {\bibinfo {volume} {118}},\ \bibinfo {pages} {240403} (\bibinfo {year}
  {2017})}\BibitemShut {NoStop}%
\bibitem [{\citenamefont {Pe\~na Ardila}\ \emph {et~al.}(2018)\citenamefont
  {Pe\~na Ardila}, \citenamefont {Heyl},\ and\ \citenamefont
  {Eckardt}}]{Ardilla2018}%
  \BibitemOpen
  \bibfield  {author} {\bibinfo {author} {\bibfnamefont {L.~A.}\ \bibnamefont
  {Pe\~na Ardila}}, \bibinfo {author} {\bibfnamefont {M.}~\bibnamefont
  {Heyl}},\ and\ \bibinfo {author} {\bibfnamefont {A.}~\bibnamefont
  {Eckardt}},\ }\bibfield  {title} {\bibinfo {title} {Measuring the
  single-particle density matrix for fermions and hard-core bosons in an
  optical lattice},\ }\href {https://doi.org/10.1103/PhysRevLett.121.260401}
  {\bibfield  {journal} {\bibinfo  {journal} {Phys. Rev. Lett.}\ }\textbf
  {\bibinfo {volume} {121}},\ \bibinfo {pages} {260401} (\bibinfo {year}
  {2018})}\BibitemShut {NoStop}%
\bibitem [{\citenamefont {Zheng}\ \emph {et~al.}(2020)\citenamefont {Zheng},
  \citenamefont {Irsigler}, \citenamefont {Jiang}, \citenamefont {Weitenberg},\
  and\ \citenamefont {Hofstetter}}]{Zheng2020}%
  \BibitemOpen
  \bibfield  {author} {\bibinfo {author} {\bibfnamefont {J.-H.}\ \bibnamefont
  {Zheng}}, \bibinfo {author} {\bibfnamefont {B.}~\bibnamefont {Irsigler}},
  \bibinfo {author} {\bibfnamefont {L.}~\bibnamefont {Jiang}}, \bibinfo
  {author} {\bibfnamefont {C.}~\bibnamefont {Weitenberg}},\ and\ \bibinfo
  {author} {\bibfnamefont {W.}~\bibnamefont {Hofstetter}},\ }\bibfield  {title}
  {\bibinfo {title} {Measuring an interaction-induced topological phase
  transition via the single-particle density matrix},\ }\href
  {https://doi.org/10.1103/PhysRevA.101.013631} {\bibfield  {journal} {\bibinfo
   {journal} {Phys. Rev. A}\ }\textbf {\bibinfo {volume} {101}},\ \bibinfo
  {pages} {013631} (\bibinfo {year} {2020})}\BibitemShut {NoStop}%
\bibitem [{\citenamefont {Hatano}\ and\ \citenamefont
  {Nelson}(1996)}]{Hatano1996}%
  \BibitemOpen
  \bibfield  {author} {\bibinfo {author} {\bibfnamefont {N.}~\bibnamefont
  {Hatano}}\ and\ \bibinfo {author} {\bibfnamefont {D.~R.}\ \bibnamefont
  {Nelson}},\ }\bibfield  {title} {\bibinfo {title} {Localization transitions
  in non-{H}ermitian quantum mechanics},\ }\href
  {https://doi.org/10.1103/PhysRevLett.77.570} {\bibfield  {journal} {\bibinfo
  {journal} {Phys. Rev. Lett.}\ }\textbf {\bibinfo {volume} {77}},\ \bibinfo
  {pages} {570} (\bibinfo {year} {1996})}\BibitemShut {NoStop}%
\bibitem [{\citenamefont {Ashida}\ \emph {et~al.}(2020)\citenamefont {Ashida},
  \citenamefont {Gong},\ and\ \citenamefont {Ueda}}]{Ashida2020}%
  \BibitemOpen
  \bibfield  {author} {\bibinfo {author} {\bibfnamefont {Y.}~\bibnamefont
  {Ashida}}, \bibinfo {author} {\bibfnamefont {Z.}~\bibnamefont {Gong}},\ and\
  \bibinfo {author} {\bibfnamefont {M.}~\bibnamefont {Ueda}},\ }\bibfield
  {title} {\bibinfo {title} {Non-{H}ermitian physics},\ }\Eprint
  {https://arxiv.org/abs/2006.01837} {arXiv:2006.01837 [cond-mat.mes-hall]}
  (\bibinfo {year} {2020})\BibitemShut {NoStop}%
\bibitem [{\citenamefont {Yang}\ and\ \citenamefont {Bhatt}(1996)}]{Yang1996}%
  \BibitemOpen
  \bibfield  {author} {\bibinfo {author} {\bibfnamefont {K.}~\bibnamefont
  {Yang}}\ and\ \bibinfo {author} {\bibfnamefont {R.~N.}\ \bibnamefont
  {Bhatt}},\ }\bibfield  {title} {\bibinfo {title} {Floating of extended states
  and localization transition in a weak magnetic field},\ }\href
  {https://www.doi.org/10.1103/physrevlett.76.1316} {\bibfield  {journal}
  {\bibinfo  {journal} {Physical Review Letters}\ }\textbf {\bibinfo {volume}
  {76}},\ \bibinfo {pages} {1316} (\bibinfo {year} {1996})}\BibitemShut
  {NoStop}%
\bibitem [{\citenamefont {Bianco}\ and\ \citenamefont
  {Resta}(2011)}]{Bianco2011}%
  \BibitemOpen
  \bibfield  {author} {\bibinfo {author} {\bibfnamefont {R.}~\bibnamefont
  {Bianco}}\ and\ \bibinfo {author} {\bibfnamefont {R.}~\bibnamefont {Resta}},\
  }\bibfield  {title} {\bibinfo {title} {Mapping topological order in
  coordinate space},\ }\href {https://www.doi.org/10.1103/physrevb.84.241106}
  {\bibfield  {journal} {\bibinfo  {journal} {Physical Review B}\ }\textbf
  {\bibinfo {volume} {84}},\ \bibinfo {pages} {241106} (\bibinfo {year}
  {2011})}\BibitemShut {NoStop}%
\bibitem [{SM()}]{SM}%
  \BibitemOpen
  \bibinfo {note} {See the supplmental material [URL] for technical
  details}\BibitemShut {NoStop}%
\bibitem [{\citenamefont {Crispin~Gardiner}(2004)}]{CrispinGardiner2004}%
  \BibitemOpen
  \bibfield  {author} {\bibinfo {author} {\bibfnamefont {P.~Z.}\ \bibnamefont
  {Crispin~Gardiner}},\ }\href
  {https://www.ebook.de/de/product/3014096/crispin_gardiner_peter_zoller_quantum_noise.html}
  {\emph {\bibinfo {title} {Quantum Noise}}}\ (\bibinfo  {publisher} {Springer
  Berlin Heidelberg},\ \bibinfo {year} {2004})\BibitemShut {NoStop}%
\bibitem [{\citenamefont {Schwarz}\ \emph {et~al.}(2016)\citenamefont
  {Schwarz}, \citenamefont {Goldstein}, \citenamefont {Dorda}, \citenamefont
  {Arrigoni}, \citenamefont {Weichselbaum},\ and\ \citenamefont {von
  Delft}}]{Schwarz2016}%
  \BibitemOpen
  \bibfield  {author} {\bibinfo {author} {\bibfnamefont {F.}~\bibnamefont
  {Schwarz}}, \bibinfo {author} {\bibfnamefont {M.}~\bibnamefont {Goldstein}},
  \bibinfo {author} {\bibfnamefont {A.}~\bibnamefont {Dorda}}, \bibinfo
  {author} {\bibfnamefont {E.}~\bibnamefont {Arrigoni}}, \bibinfo {author}
  {\bibfnamefont {A.}~\bibnamefont {Weichselbaum}},\ and\ \bibinfo {author}
  {\bibfnamefont {J.}~\bibnamefont {von Delft}},\ }\bibfield  {title} {\bibinfo
  {title} {Lindblad-driven discretized leads for nonequilibrium steady-state
  transport in quantum impurity models: Recovering the continuum limit},\
  }\href {https://doi.org/10.1103/PhysRevB.94.155142} {\bibfield  {journal}
  {\bibinfo  {journal} {Phys. Rev. B}\ }\textbf {\bibinfo {volume} {94}},\
  \bibinfo {pages} {155142} (\bibinfo {year} {2016})}\BibitemShut {NoStop}%
\bibitem [{\citenamefont {Hofstadter}(1976)}]{Hofstadter1976}%
  \BibitemOpen
  \bibfield  {author} {\bibinfo {author} {\bibfnamefont {D.~R.}\ \bibnamefont
  {Hofstadter}},\ }\bibfield  {title} {\bibinfo {title} {Energy levels and wave
  functions of {B}loch electrons in rational and irrational magnetic fields},\
  }\href {https://www.doi.org/10.1103/physrevb.14.2239} {\bibfield  {journal}
  {\bibinfo  {journal} {Physical Review B}\ }\textbf {\bibinfo {volume} {14}},\
  \bibinfo {pages} {2239} (\bibinfo {year} {1976})}\BibitemShut {NoStop}%
\bibitem [{\citenamefont {Goodman}(2020)}]{Goodman2020}%
  \BibitemOpen
  \bibfield  {author} {\bibinfo {author} {\bibfnamefont {J.}~\bibnamefont
  {Goodman}},\ }\href@noop {} {\emph {\bibinfo {title} {Speckle phenomena in
  optics : theory and applications}}}\ (\bibinfo  {publisher} {SPIE Press},\
  \bibinfo {address} {Bellingham, Washington},\ \bibinfo {year}
  {2020})\BibitemShut {NoStop}%
\bibitem [{\citenamefont {Niu}\ \emph {et~al.}(1985)\citenamefont {Niu},
  \citenamefont {Thouless},\ and\ \citenamefont {Wu}}]{Niu1985}%
  \BibitemOpen
  \bibfield  {author} {\bibinfo {author} {\bibfnamefont {Q.}~\bibnamefont
  {Niu}}, \bibinfo {author} {\bibfnamefont {D.~J.}\ \bibnamefont {Thouless}},\
  and\ \bibinfo {author} {\bibfnamefont {Y.-S.}\ \bibnamefont {Wu}},\
  }\bibfield  {title} {\bibinfo {title} {Quantized {H}all conductance as a
  topological invariant},\ }\href
  {https://www.doi.org/10.1103/physrevb.31.3372} {\bibfield  {journal}
  {\bibinfo  {journal} {Physical Review B}\ }\textbf {\bibinfo {volume} {31}},\
  \bibinfo {pages} {3372} (\bibinfo {year} {1985})}\BibitemShut {NoStop}%
\bibitem [{\citenamefont {Fukui}\ \emph {et~al.}(2005)\citenamefont {Fukui},
  \citenamefont {Hatsugai},\ and\ \citenamefont {Suzuki}}]{Fukui2005}%
  \BibitemOpen
  \bibfield  {author} {\bibinfo {author} {\bibfnamefont {T.}~\bibnamefont
  {Fukui}}, \bibinfo {author} {\bibfnamefont {Y.}~\bibnamefont {Hatsugai}},\
  and\ \bibinfo {author} {\bibfnamefont {H.}~\bibnamefont {Suzuki}},\
  }\bibfield  {title} {\bibinfo {title} {{C}hern numbers in discretized
  {B}rillouin zone: Efficient method of computing (spin) {H}all conductances},\
  }\href {https://www.doi.org/10.1143/jpsj.74.1674} {\bibfield  {journal}
  {\bibinfo  {journal} {Journal of the Physical Society of Japan}\ }\textbf
  {\bibinfo {volume} {74}},\ \bibinfo {pages} {1674} (\bibinfo {year}
  {2005})}\BibitemShut {NoStop}%
\bibitem [{\citenamefont {Caio}\ \emph {et~al.}(2019)\citenamefont {Caio},
  \citenamefont {Moller}, \citenamefont {Cooper},\ and\ \citenamefont
  {Bhaseen}}]{Caio2019}%
  \BibitemOpen
  \bibfield  {author} {\bibinfo {author} {\bibfnamefont {M.~D.}\ \bibnamefont
  {Caio}}, \bibinfo {author} {\bibfnamefont {G.}~\bibnamefont {Moller}},
  \bibinfo {author} {\bibfnamefont {N.~R.}\ \bibnamefont {Cooper}},\ and\
  \bibinfo {author} {\bibfnamefont {M.~J.}\ \bibnamefont {Bhaseen}},\
  }\bibfield  {title} {\bibinfo {title} {Topological marker currents in {C}hern
  insulators},\ }\href {https://www.doi.org/10.1038/s41567-018-0390-7}
  {\bibfield  {journal} {\bibinfo  {journal} {Nature Physics}\ }\textbf
  {\bibinfo {volume} {15}},\ \bibinfo {pages} {257} (\bibinfo {year}
  {2019})}\BibitemShut {NoStop}%
\bibitem [{\citenamefont {MacKinnon}\ and\ \citenamefont
  {Kramer}(1983)}]{MacKinnon1983}%
  \BibitemOpen
  \bibfield  {author} {\bibinfo {author} {\bibfnamefont {A.}~\bibnamefont
  {MacKinnon}}\ and\ \bibinfo {author} {\bibfnamefont {B.}~\bibnamefont
  {Kramer}},\ }\bibfield  {title} {\bibinfo {title} {The scaling theory of
  electrons in disordered solids: Additional numerical results},\ }\href
  {https://www.doi.org/10.1007/bf01578242} {\bibfield  {journal} {\bibinfo
  {journal} {Zeitschrift fur Physik B Condensed Matter}\ }\textbf {\bibinfo
  {volume} {53}},\ \bibinfo {pages} {1} (\bibinfo {year} {1983})}\BibitemShut
  {NoStop}%
\bibitem [{\citenamefont {Loring}\ and\ \citenamefont
  {Hastings}(2010)}]{Loring2010}%
  \BibitemOpen
  \bibfield  {author} {\bibinfo {author} {\bibfnamefont {T.~A.}\ \bibnamefont
  {Loring}}\ and\ \bibinfo {author} {\bibfnamefont {M.~B.}\ \bibnamefont
  {Hastings}},\ }\bibfield  {title} {\bibinfo {title} {Disordered topological
  insulators via {$\mathrm{C}^*$}-algebras},\ }\href
  {https://doi.org/10.1209/0295-5075/92/67004} {\bibfield  {journal} {\bibinfo
  {journal} {{EPL} (Europhysics Letters)}\ }\textbf {\bibinfo {volume} {92}},\
  \bibinfo {pages} {67004} (\bibinfo {year} {2010})}\BibitemShut {NoStop}%
\bibitem [{\citenamefont {Fogler}\ \emph {et~al.}(1998)\citenamefont {Fogler},
  \citenamefont {Dobin},\ and\ \citenamefont {Shklovskii}}]{Fogler1998}%
  \BibitemOpen
  \bibfield  {author} {\bibinfo {author} {\bibfnamefont {M.~M.}\ \bibnamefont
  {Fogler}}, \bibinfo {author} {\bibfnamefont {A.~Y.}\ \bibnamefont {Dobin}},\
  and\ \bibinfo {author} {\bibfnamefont {B.~I.}\ \bibnamefont {Shklovskii}},\
  }\bibfield  {title} {\bibinfo {title} {Localization length at the resistivity
  minima of the quantum hall effect},\ }\href
  {https://doi.org/10.1103/PhysRevB.57.4614} {\bibfield  {journal} {\bibinfo
  {journal} {Phys. Rev. B}\ }\textbf {\bibinfo {volume} {57}},\ \bibinfo
  {pages} {4614} (\bibinfo {year} {1998})}\BibitemShut {NoStop}%
\bibitem [{\citenamefont {Ostrovsky}\ \emph {et~al.}(2007)\citenamefont
  {Ostrovsky}, \citenamefont {Gornyi},\ and\ \citenamefont
  {Mirlin}}]{Ostrovsky2007}%
  \BibitemOpen
  \bibfield  {author} {\bibinfo {author} {\bibfnamefont {P.~M.}\ \bibnamefont
  {Ostrovsky}}, \bibinfo {author} {\bibfnamefont {I.~V.}\ \bibnamefont
  {Gornyi}},\ and\ \bibinfo {author} {\bibfnamefont {A.~D.}\ \bibnamefont
  {Mirlin}},\ }\bibfield  {title} {\bibinfo {title} {Quantum criticality and
  minimal conductivity in graphene with long-range disorder},\ }\href
  {https://doi.org/10.1103/PhysRevLett.98.256801} {\bibfield  {journal}
  {\bibinfo  {journal} {Phys. Rev. Lett.}\ }\textbf {\bibinfo {volume} {98}},\
  \bibinfo {pages} {256801} (\bibinfo {year} {2007})}\BibitemShut {NoStop}%
\bibitem [{\citenamefont {Rycerz}\ \emph {et~al.}(2007)\citenamefont {Rycerz},
  \citenamefont {Tworzyd{\l}o},\ and\ \citenamefont {Beenakker}}]{Rycerz2007}%
  \BibitemOpen
  \bibfield  {author} {\bibinfo {author} {\bibfnamefont {A.}~\bibnamefont
  {Rycerz}}, \bibinfo {author} {\bibfnamefont {J.}~\bibnamefont
  {Tworzyd{\l}o}},\ and\ \bibinfo {author} {\bibfnamefont {C.~W.~J.}\
  \bibnamefont {Beenakker}},\ }\bibfield  {title} {\bibinfo {title}
  {Anomalously large conductance fluctuations in weakly disordered graphene},\
  }\href {https://doi.org/10.1209/0295-5075/79/57003} {\bibfield  {journal}
  {\bibinfo  {journal} {Europhysics Letters ({EPL})}\ }\textbf {\bibinfo
  {volume} {79}},\ \bibinfo {pages} {57003} (\bibinfo {year}
  {2007})}\BibitemShut {NoStop}%
\bibitem [{\citenamefont {Silberstein}\ \emph {et~al.}(2020)\citenamefont
  {Silberstein}, \citenamefont {Behrends}, \citenamefont {Goldstein},\ and\
  \citenamefont {Ilan}}]{Silberstein2020}%
  \BibitemOpen
  \bibfield  {author} {\bibinfo {author} {\bibfnamefont {N.}~\bibnamefont
  {Silberstein}}, \bibinfo {author} {\bibfnamefont {J.}~\bibnamefont
  {Behrends}}, \bibinfo {author} {\bibfnamefont {M.}~\bibnamefont
  {Goldstein}},\ and\ \bibinfo {author} {\bibfnamefont {R.}~\bibnamefont
  {Ilan}},\ }\bibfield  {title} {\bibinfo {title} {Berry connection induced
  anomalous wave-packet dynamics in non-{H}ermitian systems},\ }\Eprint
  {https://arxiv.org/abs/2004.13746} {arXiv:2004.13746 [cond-mat.mes-hall]}
  (\bibinfo {year} {2020})\BibitemShut {NoStop}%
\bibitem [{\citenamefont {Nandkishore}\ and\ \citenamefont
  {Huse}(2015)}]{Nandkishore2015}%
  \BibitemOpen
  \bibfield  {author} {\bibinfo {author} {\bibfnamefont {R.}~\bibnamefont
  {Nandkishore}}\ and\ \bibinfo {author} {\bibfnamefont {D.~A.}\ \bibnamefont
  {Huse}},\ }\bibfield  {title} {\bibinfo {title} {Many-body localization and
  thermalization in quantum statistical mechanics},\ }\href
  {https://doi.org/10.1146/annurev-conmatphys-031214-014726} {\bibfield
  {journal} {\bibinfo  {journal} {Annual Review of Condensed Matter Physics}\
  }\textbf {\bibinfo {volume} {6}},\ \bibinfo {pages} {15} (\bibinfo {year}
  {2015})}\BibitemShut {NoStop}%
\bibitem [{\citenamefont {Altman}\ and\ \citenamefont
  {Vosk}(2015)}]{Altman2015}%
  \BibitemOpen
  \bibfield  {author} {\bibinfo {author} {\bibfnamefont {E.}~\bibnamefont
  {Altman}}\ and\ \bibinfo {author} {\bibfnamefont {R.}~\bibnamefont {Vosk}},\
  }\bibfield  {title} {\bibinfo {title} {Universal dynamics and renormalization
  in many-body-localized systems},\ }\href
  {https://doi.org/10.1146/annurev-conmatphys-031214-014701} {\bibfield
  {journal} {\bibinfo  {journal} {Annual Review of Condensed Matter Physics}\
  }\textbf {\bibinfo {volume} {6}},\ \bibinfo {pages} {383} (\bibinfo {year}
  {2015})}\BibitemShut {NoStop}%
\end{thebibliography}%

\clearpage

\setcounter{equation}{0}
\setcounter{figure}{0}
\setcounter{table}{0}
\setcounter{page}{1}
\makeatletter
\renewcommand{\theequation}{S\arabic{equation}}
\renewcommand{\thefigure}{SF\arabic{figure}}
\renewcommand{\thetable}{ST\arabic{table}}
\renewcommand{\thesection}{S.\Roman{section}}

\begin{widetext}

\section*{Supplemental Material for: ``Disorder in dissipation-induced topological states: Evidence for a different type of localization transition''}

In this Supplemental Material we provide additional technical details and results.
In Sec.~\ref{sec:href_disorder} we display an argument to the fact that when the disorder appears in the system-bath coupling Hamiltonian, the localization phase transition is in the same universality class as in equilibrium.
In Secs.~\ref{sec:methodI}--\ref{sec:methodIII} we present additional details regarding the three methods which were used to calculate the critical exponent $\nu$ out of equilibrium.
\red{In Sec.~\ref{sec:universality} we discuss the choice of parameters and verify that the critical exponent is universal, i.e., independent of the exact parameter values.}
Finally, in Sec.~\ref{sec:ginv} we characterize the distribution and correlation of the elements of the matrix $G^{-1}$.

\section{\label{subsec:Disorder-in-the}Disorder in the dissipative dynamics}
\label{sec:href_disorder}

In the main text we have stated that if we consider disorder only
in the reference Hamiltonian $h^{\mathrm{ref}}$ (that is, the dynamics
is purely-dissipative, $H_S=0$), then the critical exponent of the phase transition
is the same as in equilibrium (that is, in the same universality class).
We now present a more detailed argument for this. In fact, it is a
special case of the following claim: 

\textit{Claim.} Let $h$ be a nondegenerate Hamiltonian with a property $\xi_{h}(E)$,
which is a function of the eigenstates of $h$ (with energy $E$).
Suppose we have a phase transition described by a scaling law, $\xi_{h}(E)\propto\left|E-E_{c}\right|^{-\nu}$,
where $E_{c}$ is the critical energy and $\nu$ is the critical exponent.
Then, any analytic function $G=g(h)$ of $h$ that satisfies
\begin{equation}
0<\left.\frac{dg}{dh}\right|_{E_{c}}<\infty,\label{eq:dg_dh}
\end{equation}
will display a phase transition with the same critical exponent. That
is, $\xi_{G}(n)\propto\left|n-n_{c}\right|^{-\nu}$, where $n$ represent
an eigenvalue of $G$.

\textit{Proof.} We notice that $G$ has the same eigenstates as $h$,
but with different eigenvalues described by the relation $n(E)=g(E)$,
where $n(E)$ is the eigenvalue of $G$ corresponding to the eigenvalue
$E$ of $h$. Since $\xi$ is determined only by the eigenstates,
we have
\begin{equation}
\xi_{G}(n(E))=\xi_{h}(E),
\end{equation}
and thus
\begin{align}
\xi_{G}(n)\propto & =|E-E_{c}|^{-\nu}=|g^{-1}(n)-g^{-1}(n_{c})|^{-\nu},
\end{align}
where $g^{-1}$ is the inverse of the function $g$. We note that $g^{-1}$ is
well defined around $n_{c}$ since $0<\left.\frac{dg}{dh}\right|_{E_{c}}<\infty$.
Expanding $g^{-1}$ to first order around $n_{c}$, we get
\begin{equation}
\xi_{G}(n)\approx\left|\left.\frac{dg^{-1}}{dn}\right|_{n_{c}}(n-n_{c})\right|^{-\nu},
\end{equation}
hence $\xi(n)\propto\left|n-n_{c}\right|^{-\nu}$, as expected.

Going back to our case, 
Eq.~(5) of the main text shows that for $H_S=0$ the single-particle density matrix $G$ can be expressed as a function of $h^{\mathrm{ref}}$,
whose derivative is
\begin{equation}
\frac{dG}{dh^{\mathrm{ref}}}=-G^{2}\frac{2\pi\nu_0}{\gamma^{\mathrm{in}}}2(h^{\mathrm{ref}}-\mu^{\mathrm{eff}}),
\end{equation}
that is, condition (\ref{eq:dg_dh}) will be satisfied for $\mu^{\mathrm{eff}}\neq E_{c}$.

\section{Additional details for Method I}
\label{sec:methodI}

\textit{Calculation of the Chern number.} For efficient calculation
of the Chern number we have followed the method of 
Ref.~\cite{Fukui2005}.
We divide the parameter space $0\le\theta_{x},\theta_{y}\le2\pi$
into a grid of size $N_{g} \times N_{g}$ with equal spacing. For
each point,
\begin{equation}
\boldsymbol{\theta}=(\theta_{x},\theta_{y})=\dfrac{2\pi}{N_{g}}(r_{x},r_{y}),\qquad r_{x},r_{y}=0,\cdots,N_{g}-1,\label{eq:Ng_for_Fukui}
\end{equation}
 we define:
\begin{equation}
U_{\hat{\mu}}(\boldsymbol{\theta})\equiv\left\langle \psi(\boldsymbol{\theta})|\psi(\boldsymbol{\theta}+\hat{\mu})\right\rangle /N_{\hat{\mu}}(\boldsymbol{\theta}),\label{eq:Fukui_ground_state_U_theta}
\end{equation}
where $N_{\hat{\mu}}(\boldsymbol{\theta})\equiv\left|\left\langle \psi(\boldsymbol{\theta})|\psi(\boldsymbol{\theta}+\hat{\mu})\right\rangle \right|$
is a normalization factor, and $\hat{\mu}=\hat{x},\hat{y}$ is a grid lattice
vector in the $x$ or $y$ direction, respectively.
We define a discretized version of the Berry curvature,
\begin{equation}
F(\boldsymbol{\theta})\equiv\ln \left[ U_{x}(\boldsymbol{\theta})U_{y}(\boldsymbol{\theta}+\hat{x})U_{x}(\boldsymbol{\theta}+\hat{y})^{-1}U_{y}(\boldsymbol{\theta})^{-1} \right],\label{eq:discrete_curvature}
\end{equation}
where the principal branch of the logarithm is used. 
Finally, the Chern
number is defined as
\begin{equation}
C\equiv\dfrac{1}{2\pi i}\sum_{\boldsymbol{\theta}}F(\boldsymbol{\theta}),
\end{equation}
which must result in an integer value. However, the result might
contain an error if $N_{g}$ is not large enough. In our case we can detect errors
by checking that the sum of the Chern numbers of all the single-particle
states equals $-1$ (the total Chern number of the first Landau band).
If the result is different than $-1$, we know that at least one error
has occurred in the calculation and therefore reject it. We have
taken values of $N_{g}$ which would keep the rejection rate smaller
than 2\%: In equilibrium, we choose $N_{g}=30$ for all of
the system sizes. Out of equilibrium, we choose $N_{g}=25$ for $L=7,...,49$,
and $N_{g}=31$ for $L=56,63$.%

\begin{table}
	\begin{centering}
		\begin{tabular}{|c|c|c|c|c|c|c|}
			\multicolumn{3}{c}{Equilibrium} & \multicolumn{1}{c}{} & \multicolumn{3}{c}{Out of equilibrium}\tabularnewline
			\cline{1-3} \cline{5-7} 
			~$L_{\mathrm{min}}$~ & $\nu$ & ~$\chi_{\mathrm{red}}^{2}$~ &  & ~$L_{\mathrm{min}}$~ & $\nu$ & ~$\chi_{\mathrm{red}}^{2}$~\tabularnewline
			\cline{1-3} \cline{5-7} 
			7 & ~$2.36\pm0.01$~ & 46.4 &  & 7 & ~$3.00\pm0.01$~ & 18\tabularnewline
			\cline{1-3} \cline{5-7} 
			14 & $2.61\pm0.02$ & 0.92 &  & 14 & $3.26\pm0.03$ & 1.2\tabularnewline
			\cline{1-3} \cline{5-7} 
			21 & $2.63\pm0.03$ & 0.94 &  & 21 & $3.22\pm0.05$ & 1.2\tabularnewline
			\cline{1-3} \cline{5-7} 
			28 & $2.58\pm0.04$ & 0.57 &  & 28 & $3.15\pm0.06$ & 1.0\tabularnewline
			\cline{1-3} \cline{5-7} 
			35 & $2.64\pm0.07$ & 0.32 &  & 35 & $2.99\pm0.10$ & 0.28\tabularnewline
			\cline{1-3} \cline{5-7} 
			42 & $2.62\pm0.12$ & 0.46 & ~~~~~~~~~ & 42 & $2.97\pm0.19$ & 0.41\tabularnewline
			\cline{1-3} \cline{5-7} 
		\end{tabular} 
		\par\end{centering}
	\caption{\label{tab:coundcting_states_nu_results}The results of the critical
		exponent $\nu$ without correction to scaling, extracted from fits
		of the number of conducting states, Eq.~(\ref{eqn:nc}). The smallest system size is taken
		as $L_{\mathrm{min}}$ and the largest system size is always $L_{\mathrm{max}}=63$.}
\end{table}

\textit{Calculation of the critical exponent.}
Unlike an infinite system, in which an extended (conducting) state exists only at a single
energy $E_{c}$, in a finite-sized system there is a range of extended
states, corresponding to the range of energies 
with localization lengths $\xi(E) > L$. 
Thus, to obtain the finite-size mobility edges need to solve $\xi(E)=\xi_{0}\left|E-E_{c}\right|^{-\nu}=L$,
leading to $E_{1,2}=E_{c}\pm (\xi_{0}/L)^{1/\nu}.$
Therefore, the number of conducting states would be $N_{c}=L^{2}\int_{E_{1}}^{E_{2}}\rho(E)dE,$
where $\rho(E)$ is the density of states of the band. Approximating
$\rho(E)\thickapprox\rho(E_{c})$ and recalling that the total number of states in a Landau band is $N_{b}=\alpha L^{2}$ ($\alpha=1/7$), 
we obtain the scaling relation
\begin{equation} \label{eqn:nc}
\dfrac{N_{c}}{N_{b}}=aL^{-1/\nu},
\end{equation}
where $a$ is some constant. $\nu$ can be extracted by numerical
calculations of $N_{c}$ for different system sizes. The results (without
corrections to scaling) in and out of equilibrium are presented in
Table~\ref{tab:coundcting_states_nu_results}. In equilibrium, the
corrections to scaling are significant only when $L_{\mathrm{min}}=7$.
The first correction can be included by considering the generalized
scaling form:
\begin{equation}
\dfrac{N_{c}}{N_{b}}=a\left(1+bL^{-y}\right)L^{-1/\nu},\label{eq:conducting_states_correction}
\end{equation}
where $y>0$ is the leading irrelevant exponent. We managed to consistently
include corrections to scaling only when the lowest system size is
included (that is, $L=7,...,63$). This leads to $\nu=2.63\pm0.02$,
$y=4.6\pm0.71$, $\chi_{\mathrm{red}}^{2}=0.99$, which is in agreement
with the values obtained without corrections to scaling but with
the lower system sizes being excluded (as described below).
Out of equilibrium, a single correction
to scaling in the form of Eq.~(\ref{eq:conducting_states_correction})
is not compatible with the data even when the lowest system size is
included. Therefore, we base our final result only on fits without
corrections to scaling with the lower system sizes excluded.
The lowest included system size was chosen as $L_\mathrm{min}=28$ in equilibrium and  $L_\mathrm{min}=35$ out of equilibrium. This choice ensures that the three lowest system sizes are excluded to avoid corrections to scaling, but also that the change in $\nu$ between the employed $L_\mathrm{min}$ and using the following value $L_\mathrm{min}+1/\alpha = L_\mathrm{min}+7$ is smaller than the uncertainty in $\nu$.

An additional way for extracting the critical exponent is by looking at the
width of the density of the conducting states, $\rho_c(E)$, defined as
\begin{equation} \label{eqn:rhocwidth}
	\Delta E_c^2 = \frac{L^2}{N_c} \int \rho_c(E) E^2 dE - \left[ \frac{L^2}{N_c} \int \rho_c(E) E dE \right]^2,
\end{equation}
which is expected to scale as $\Delta E_c \sim L^{-1/\nu}$.
The results without corrections
are presented in Table~\ref{tab:coundcting_states_width_nu_results}.
While they also suggest a higher value of the critical exponent
out of equilibrium, they seem to be less reliable than the results
with the number of conducting states, as evidenced by the significantly large chi-squared values.
This may imply that the width of the distribution is more sensitive to finite-size corrections than
the number of conducting states (See the discussion around Fig.~7 of Ref.~\cite{Zhu2019}).
\begin{table}
\begin{centering}
\begin{tabular}{|c|c|c|c|c|c|c|}
\multicolumn{3}{c}{Equilibrium} & \multicolumn{1}{c}{} & \multicolumn{3}{c}{Out of equilibrium}\tabularnewline
\cline{1-3} \cline{5-7} 
~$L_{\mathrm{min}}$~ & $\nu$ & ~$\chi_{\mathrm{red}}^{2}$~ &  & ~$L_{\mathrm{min}}$~ & $\nu$ & ~$\chi_{\mathrm{red}}^{2}$~\tabularnewline
\cline{1-3} \cline{5-7} 
7 & $2.410\pm0.007$ & 3.5 &  & 7 & $3.65\pm0.01$ & 25\tabularnewline
\cline{1-3} \cline{5-7} 
14 & $2.42\pm0.01$ & 3.2 &  & 14 & $3.91\pm0.03$ & 1.7\tabularnewline
\cline{1-3} \cline{5-7} 
21 & $2.41\pm0.01$ & 3.8 &   & 21 & $3.86\pm0.04$ & 1.8\tabularnewline
\cline{1-3} \cline{5-7} 
28 & $2.39\pm0.02$ & 4.4 &  & 28 & $3.77\pm0.05$ & 0.67\tabularnewline
\cline{1-3} \cline{5-7} 
35 & $2.43\pm0.03$ & 5.1 &  & 35 & $3.68\pm0.09$ & 0.45\tabularnewline
\cline{1-3} \cline{5-7} 
42 & $2.52\pm0.06$ & 6.4 & ~~~~~~~~~ & 42 & $3.57\pm0.15$ & 0.34\tabularnewline
\cline{1-3} \cline{5-7} 
\end{tabular} 
\par\end{centering}
\caption{\label{tab:coundcting_states_width_nu_results}
The results of the
critical exponent $\nu$ without correction to scaling, extracted
from fits of the width of conducting states density, $\Delta E_c$ [cf.~Eq.~(\ref{eqn:rhocwidth})]. The smallest system size
is taken as $L_{\mathrm{min}}$ and the largest system size is always
$L_{\mathrm{max}}=63$.}
\end{table}

\section{Additional details for Method II}
\label{sec:methodII}

The results of the local Chern marker calculations are presented in Fig. \ref{fig:local_chern_full_results}.
Corrections to scaling are present at low system sizes. This can be seen in the insets, which show that the low systems sizes curves cross the large system curves away from $E_c$.
The corrections to scaling for small system sizes turn out be be difficult to fit accurately. However, they also decrease
rapidly with increasing system sizes. We therefore resort to omitting smaller system sizes and ignoring corrections to scaling. 
We note that the result also depends on the range of
energies (or occupations out of equilibrium) that were included in the fit. Therefore, we chose a range of energies (occupations) in which
the result is most stable (least sensitive to increasing or decreasing the number of included points). For example, for $L_{\mathrm{min}}=35$
we have estimated $\nu=2.26\pm0.04$ from the results presented in
Fig.~\ref{fig:choosing_nu_value_equilibrium}(a).
As for the degree $D$ of the polynomial $f(x)=\sum_{q=0}^{D}a_{q}x^{q}$ that was used to approximate the scaling function $f$,
we have verified that it is large enough
to capture the behavior in the given range, but is not too large,
so as to prevent overfitting. We have thus used $D=5$ in equilibrium and $D=7$ out of equilibrium.
The results in and out of equilibrium are shown in Table~\ref{tab:local_chern_nu_results}.
In order to avoid correction to scaling effects, we have chosen to exclude the four first system sizes. That is, the lowest included system size was chosen as $L_\mathrm{min}=35$ in and out of equilibrium. As in method I, we also verified that the change in $\nu$ with respect to using the following value $L_\mathrm{min}+1/\alpha = L_\mathrm{min}+7$ is smaller than the uncertainty in $\nu$.

\begin{figure}
\includegraphics[scale=0.5]{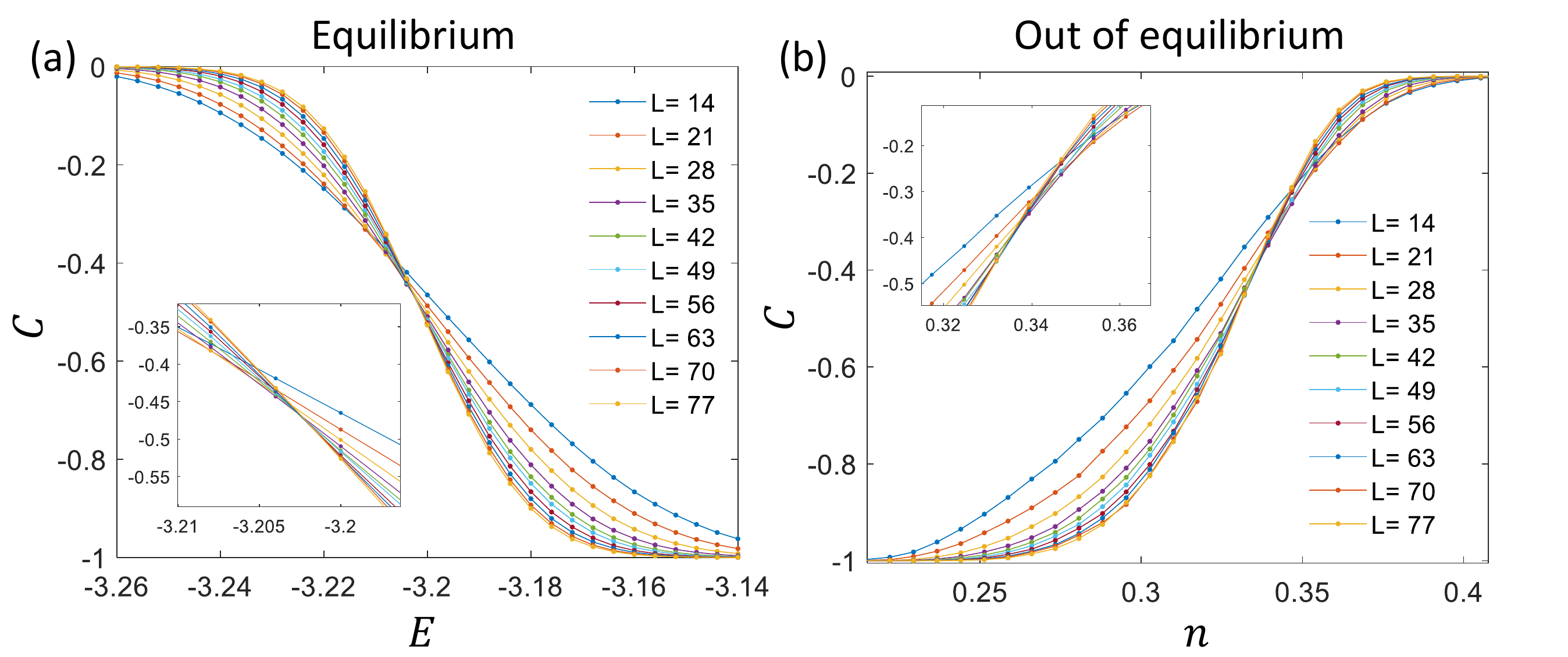}
\caption{\label{fig:local_chern_full_results}
The local Chern marker for system
sizes $L=14,...,77$, (a) in equilibrium (results averaged over $3\cdot10^{4}$ disorder realizations); (b)
out of equilibrium (results averaged over $3\cdot10^{3}$ disorder realizations).
Inset: zoom-in onto the vicinity of the critical point.
\red{We note that the data presented here includes all the data from Fig.~2 in the main text, and in addition data for smaller system sizes. This allows to see more clearly the existence of corrections to scaling in the lower system sizes.}}
\end{figure}

\begin{figure}[b]
\centering{}
\includegraphics[scale=0.5]{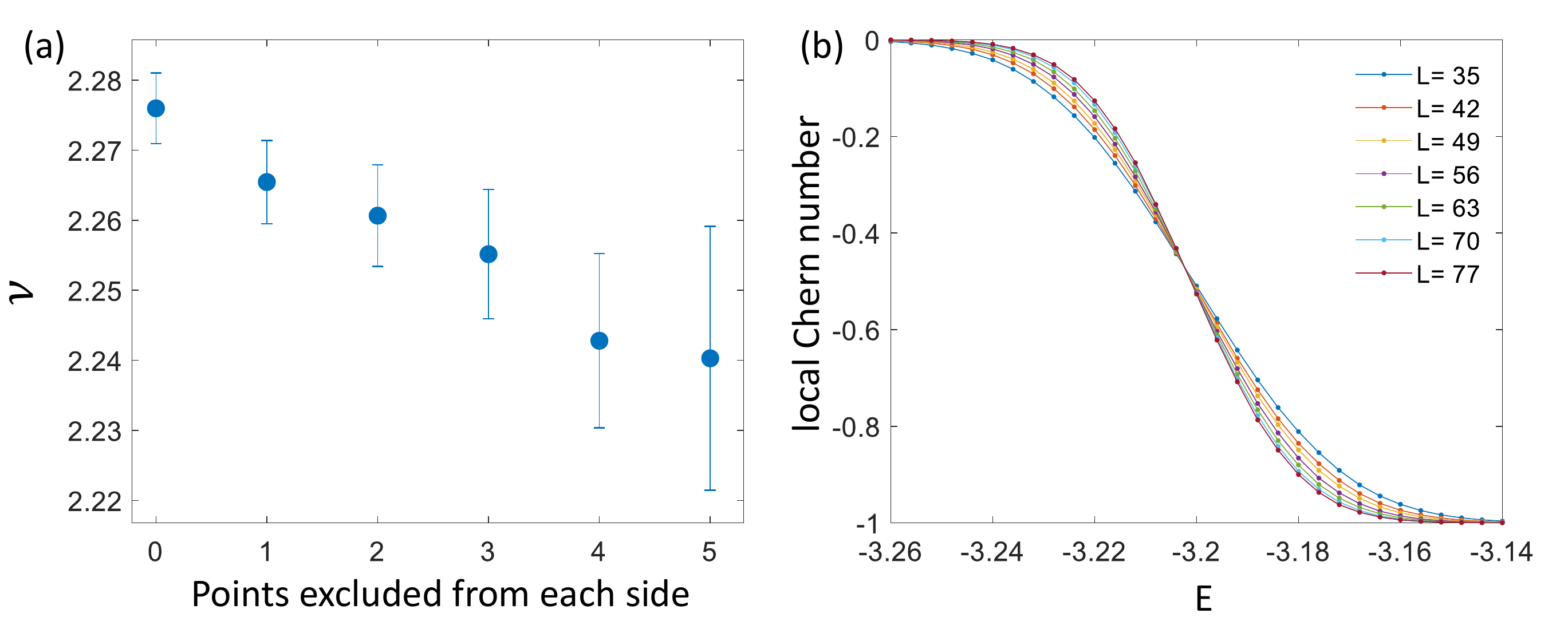}
\caption{\label{fig:choosing_nu_value_equilibrium}
Determination of $\nu$ using method II in equilibrium:
(a) Values of $\nu$ extracted
from system sizes $L=35,42,...,77$. The values on the horizontal axis denote the number
of points that were excluded from both sides of $E_c$ for each data set in panel
(b). The critical energy was found to be $E_{c} \approx -3.2025$.}
\end{figure}

\begin{table}[b]
\begin{centering}
\begin{tabular}{|c|c|c|c|c|c|c|}
\multicolumn{3}{c}{Equilibrium} & \multicolumn{1}{c}{} & \multicolumn{3}{c}{Out of equilibrium}\tabularnewline
\cline{1-3} \cline{5-7} 
~$L_{\mathrm{min}}$~ & $\nu$ & ~$\chi_{\mathrm{red}}^{2}$~ &  & ~$L_{\mathrm{min}}$~ & $\nu$ & ~$\chi_{\mathrm{red}}^{2}$~\tabularnewline
\cline{1-3} \cline{5-7} 
14 & $2.28\pm0.06$ & $40.2$ &  & 14 & $2.81\pm0.04$ & 25\tabularnewline
\cline{1-3} \cline{5-7} 
21 & $2.23\pm0.05$ & $17.4$ &  & 21 & $2.79\pm0.06$ & 12\tabularnewline
\cline{1-3} \cline{5-7} 
28 & $2.24\pm0.05$ & $8.3$ &   & 28 & $2.87\pm0.07$ & 5.4\tabularnewline
\cline{1-3} \cline{5-7} 
35 & $2.26\pm0.04$ & $5.6$ &  & 35 & $2.91\pm0.06$ & 3.5\tabularnewline
\cline{1-3} \cline{5-7} 
42 & $2.26\pm0.04$ & $6.1$ &  & 42 & $2.92\pm0.08$ & 4\tabularnewline
\cline{1-3} \cline{5-7} 
49 & $2.20\pm0.05$ & $5.3$ & ~~~~~~~~~ & 49 & $2.89\pm0.10$ & 4\tabularnewline
\cline{1-3} \cline{5-7} 
\end{tabular} 
\par\end{centering}
\caption{\label{tab:local_chern_nu_results}The results of the critical exponent
$\nu$, extracted from method II. The smallest system size is taken
as $L_\mathrm{min}$ and the largest system size is always $L_\mathrm{max}=77$.}
\end{table}

To verify these results, we have also extracted the critical exponent
from the $L$-dependence of the derivative of $C$ at the critical point. For example, in equilibrium, since $C_{L}(E)=f\left((E-E_{c})L^{\frac{1}{\nu}}\right)$, we
have 
\begin{equation}
\ln\left(\left.\frac{\partial C_{L}(E)}{\partial E}\right|_{E_{c}}\right)=\ln\left(f'(0)\right)+\frac{1}{\nu}\ln(L).
\end{equation}
By fitting a polynomial expansion to each $C_L(E)$ data set at fixed $L$, we can obtain the left hand side of the last equation by approximating
$\left.\frac{\partial C_{L}(E)}{\partial E}\right|_{E_{c}}\approx\mathrm{max}\left(\frac{\partial C_{L}(E)}{\partial E}\right)$.
Then, $\nu$ can be extracted from a linear fit of the logarithm of the latter quantity as function of $\ln(L)$. While we found this method to be less stable, its results were still in agreement with the chi-square
minimization results: In equilibrium we got $\nu \approx 2.2-2.3$,
while out of equilibrium we got $\nu \approx 2.8-3.1$.

\section{Additional details for Method III}
\label{sec:methodIII}

\textit{Equilibrium.}
\red{As we mentioned in the main text, in Method III corrections to scaling need to be taken into account, as can be seen from the $L$ dependence of the minima in Fig.~\ref{fig:zoom_localization_length}.
We note that unlike the two previous methods, corrections to scaling are significant even for larger systems sizes. Therefore, in order to obtain a good estimation for the critical exponent one should include irrelevant exponents in the scaling procedure}.
We will now present additional details regarding
the corrections to scaling that were used in the fitting procedure
in equilibrium. We assume the existence of only one irrelevant exponent
(including more than a single irrelevant exponent would on the one hand be a numerical
challenge which in general requires data with much lower uncertainties, and on the other hand seems not to be required in practice for the system sizes used).
That is, we assume the following scaling form:
\begin{equation}
\frac{L}{\xi(E)}=f(u_{r}L^{1/\nu},u_{i}L^{-y}),
\end{equation}
where $\xi(E)$ is the localization length, $f$ is some scaling function,
$u_{r},u_{i}$ are the relevant and irrelevant scaling fields, respectively,
and $y>0$ is the irrelevant exponent. We can expand the scaling fields
in the vicinity of $E_{c}$ as $u_{r}(E-E_{c})=\sum_{n=1}^{m_{r}}a_{n}(E-E_{c})^{n}$,
$u_{i}(E-E_{c})=\sum_{n=0}^{m_{i}}b_{n}(E-E_{c})^{n}$ (the term $n=0$ is absent for
the relevant field since it must vanish at the
critical point). In addition, we expand $f$ to the first order in
the irrelevant field:
\begin{equation}
f\left(u_{r}L^{1/\nu},u_{i}L^{-y}\right) \approx f_{0}\left(u_{r}L^{1/\nu}\right)+u_{i}L^{-y}f_{1}\left(u_{r}L^{1/\nu}\right),
\end{equation}
where $f_{0},f_{1}$ are some single-parameter functions. We will
now present two approaches which lead to similar results:

(i) We assume a simple form of the scaling fields: $u_{r}(E)=E-E_{c}$,
$u_{i}(E)=1$, and take $f_{0},f_{1}$ as the following polynomials:
$f_{0}(x)=\sum_{n=0}^{n_{1}}a_{n}x^{n}$, $f_{1}(x)=\sum_{n=0}^{n_{2}}b_{n}x^{n}$,
with $n_{1}=5,n_{2}=4$.

(ii) Following Ref.~\cite{Puschmann2019}, we consider
only even terms in the scaling functions: $f_{0}(x)=\sum_{n=0}^{3}a_{n}x^{2n},$
$f_{1}(x)=\sum_{n=0}^{2}b_{n}x^{2n}$, where $n_{1}=3$, $n_{2}=2$.
However, we include additional terms in the expansion of the scaling
fields: $u_{r}(E)=\sum_{n=1}^{m_{1}}c_{n}(E-E_{c})^{n}$, $u_{i}(E)=\sum_{n=0}^{m_{2}}d_{n}(E-E_{c})^{n}$,
with $m_{1}=3$, $m_{2}=1$. This can be motivated by the fact that
our data is close to being an even function around $E_{c}$, and the
small asymmetry is reflected by the odd terms of the expansion of
the scaling fields.

We have found both approaches to have low sensitivity to the choice of the smallest
system size to be included in the fit, but are still somewhat affected
by the range of energies taken around the critical energy $E_{c}$.
As in method II, we chose a range of energies in which
the result is most stable (least sensitive to increasing or decreasing the number of included points).
We also verified that increasing $n_{1},n_{2},m_{1},m_{2}$ has small
impact on the results. A comparison of the results is presented in
Fig. \ref{fig:approaches_for_corrections}. We have found approach
(b) to be slightly more stable.

\begin{figure}
	\begin{centering}
		\includegraphics[scale=0.5]{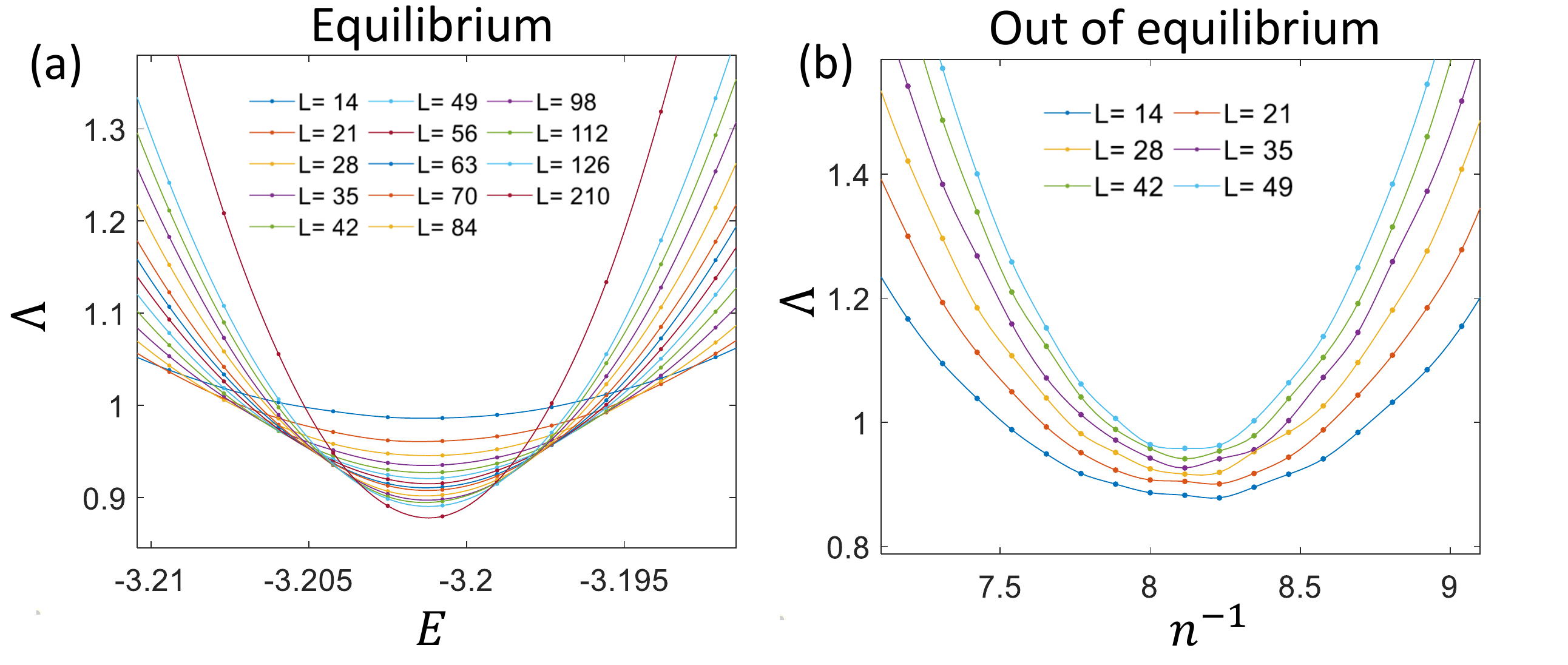}
		\par\end{centering}
	\caption{\label{fig:zoom_localization_length}
		\red{
			Dimensionless Lyapunov exponent (a) in and (b) out of equilibrium. The data presented here is the same as in Fig. 3 in the main text, but here we zoom in into the vicinity of $E_c$ ($n_{c}^{-1}$). The need for corrections to scaling (even for the larger system sizes and even in equilibrium) is clearly evidenced by the variation of the minimal value of $\Lambda$ with $L$.
		}
	}
\end{figure}

\begin{figure}
\begin{centering}
\includegraphics[scale=0.5]{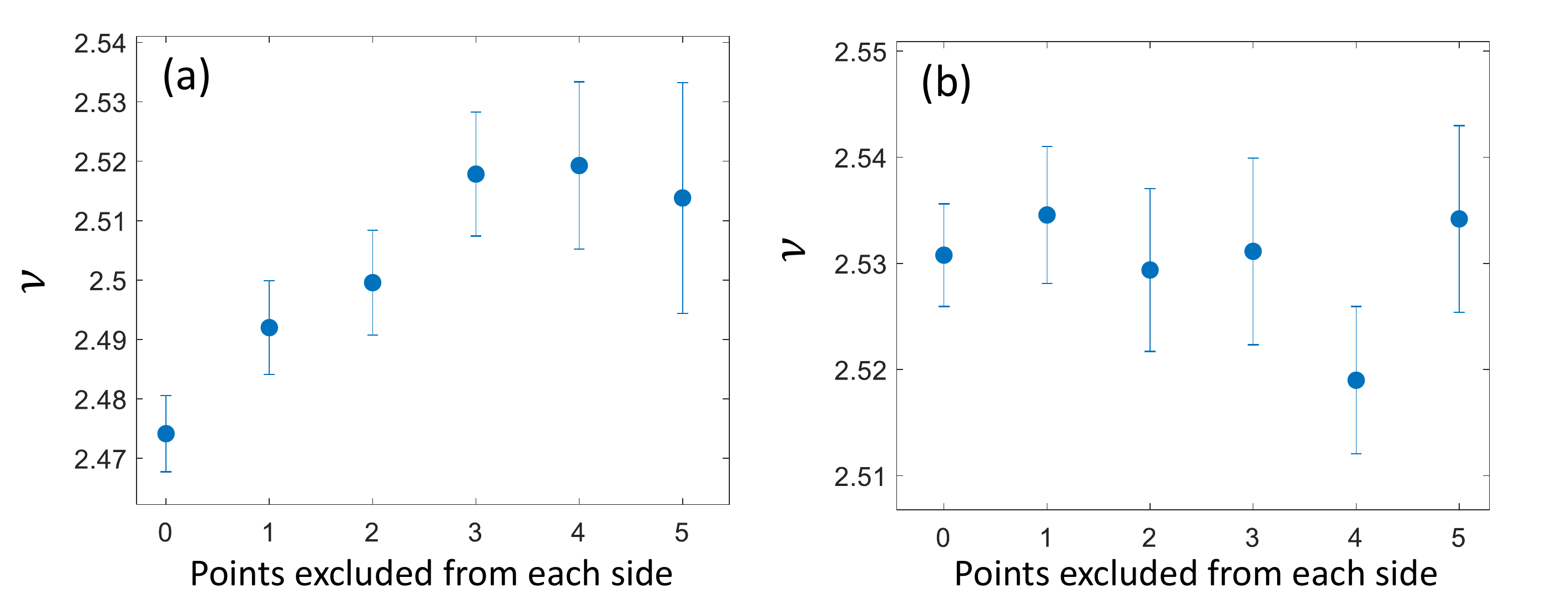}
\par\end{centering}
\caption{\label{fig:approaches_for_corrections}
Values of $\nu$ extracted in equilibrium using method III and system sizes $L=42,49,56,63,70,84,98,112,126,210$.
The values on the horizontal axis denote the number
of points that were excluded from both sides of $E_c$ for each data set in Fig.~3(a) of the main text.
Panels (a) and (b) correspond, respectively, to approaches (a) and (b) described in the text.}
\end{figure}

\textit{Out of equilibrium.} We will provide here additional details
regarding the transfer-matrix method and the choice of the parameters
that were used to extract the localization length out of equilibrium.
As described in the main text, we have calculated the matrix $G^{-1}$
(of size $L_{x}\times L$, with periodic boundary conditions in both directions), for $M$ different disorder realizations.
We then set the hopping terms in $G^{-1}$ with range along the $x$-direction larger than a cutoff
$p$ to be zero. Since the matrix now has only a \textit{finite}
hopping range, it is possible to extract the transfer-matrices by
a straightforward generalization of the conventional nearest-neighbor
case~\cite{MacKinnon1983}: Suppose that we have a $L_{x}\times L$
Hamiltonian in a quasi-1D geometry with hopping range $p$ in the
$x$-direction. It can be written as
$H=\sum_{r_x=1}^{L_{x}}\sum_{q=-p}^{p}H_{r_x,r_x+q}$,
where
$H_{r_x,r_x+q}$ is the Hamiltonian describing hopping from slab $r_x$ to slab $r_x+q$. We also define $\psi_{r_x}$ as a length-$L$ vector containing the wavefunction  amplitudes associated with the $r_x$th slab. The eigenvalue equation can
be written as: %
$\sum_{q=-p}^{p}H_{r_x,r_x+q} \psi_{r_x+q}=E \psi_{r_x}$ for each
$r_x$, where $E$ is the energy. We can isolate $\psi_{r_x+p}$
and arrive to a recursion relation
\begin{equation}
\psi_{r_x+p}=(H_{r_x,r_x+p})^{-1}\cdot\left(E \psi_{r_x}-\sum_{q=-p}^{p-1}H_{r_x,r_x+q} \psi_{r_x+q}\right).\label{eq:TMM_recursive_relation}
\end{equation}
We can then define the transfer matrix $T_{r_x}$ (of size $2pL\times2pL$)
as:
\begin{equation}
\left(\begin{array}{c}
\psi_{r_x+p}\\
\vdots\\
\psi_{r_x-p+1}
\end{array}\right)=\left(\begin{array}{ccccc}
\cdots & \cdots & (\ref{eq:TMM_recursive_relation}) & \cdots & \cdots\\
\mathbb{I} & 0 & \cdots & 0 & 0\\
0 & \mathbb{I} & \cdots & 0 & 0\\
\vdots & \vdots & \ddots & \vdots & \vdots\\
0 & 0 & \cdots & \mathbb{I} & 0
\end{array}\right)\left(\begin{array}{c}
\psi_{r_x+p-1}\\
\vdots\\
\psi_{r_x-p}
\end{array}\right)\equiv T_{r_x}\left(\begin{array}{c}
\psi_{r_x+p-1}\\
\vdots\\
\psi_{r_x-p}
\end{array}\right),\label{eq:transfer_matrix}
\end{equation}
where in the first line appear the corresponding components of equation
(\ref{eq:TMM_recursive_relation}), and $\mathbb{I}$ is the $L\times L$
identity matrix.

Therefore, from each $G^{-1}$ matrix we can extract $K=L_{x}-2c$
transfer matrices, where $c=7$ is a ``safety margin'',
which was chosen to be larger than $p$ in order to avoid mixing between
the first and the last transfer matrices of $G^{-1}$. The
effective length would then be $L_{\mathrm{eff}}=MK$. As for the choice of values for $p$ and $L_{x}$,
a priori it seems that the bigger $L_{x}$ and $p$ are, the smaller
the resulting error (since the approximation becomes more accurate).
While this is indeed the case for $L_{x}$, for $p$ the situation is more
subtle: To use Eq.~(\ref{eq:TMM_recursive_relation}) we are
required to calculate the inverse of $H_{r_{x},r_{x}+p}$, which become exponentially
small as $p$ becomes larger. Therefore, $p$ that is too large will
lead to large numerical uncertainties in the inverse matrix.

In order to examine the effects of the value $L_{x}$ on the calculation, we
first set $p=2$ [which is the exact hopping range of $G^{-1}$ for
the case of no disorder in the system Hamiltonian, see 
Eq.~(5) of the main text]
and investigate the dimensionless Lyapunov exponent $\Lambda$ for
different $L_{x}$, see Fig.~\ref{fig:Lx_p}(a)--(c).
We can see that $L_{x}=105$ is already close to the limiting value.
Then, we set $L_{x}=105$ and investigate $\Lambda$ for different
values of $p$, see Fig.~\ref{fig:Lx_p}(d)-(f). In
addition, we have performed a calculation of the critical exponent
for $L_{x}=105$ and several values of $p$, and verified that
$p=4$ and $p=5$ already give similar results. Based on this information, we chose
$L_{x}=105$ and $p=5$ in the calculations that are presented in
the main text.

\red{As can be seen in Fig.~\ref{fig:zoom_localization_length}(b), corrections to scaling are needed to be accounted for also in the nonequilibrium case. However, since the effective $L_x$ out of equilibrium is about 100-fold smaller than $L_x$ in equilibrium (see Table I in the main text), the errors are about 10 times larger than in equilibrium. This prevents us from reliably including corrections to scaling in our fits. 
Therefore, we resolved to use the approach described in the main text, see in particular Fig.~3(d) there.
}

\begin{figure}
\includegraphics[scale=0.5]{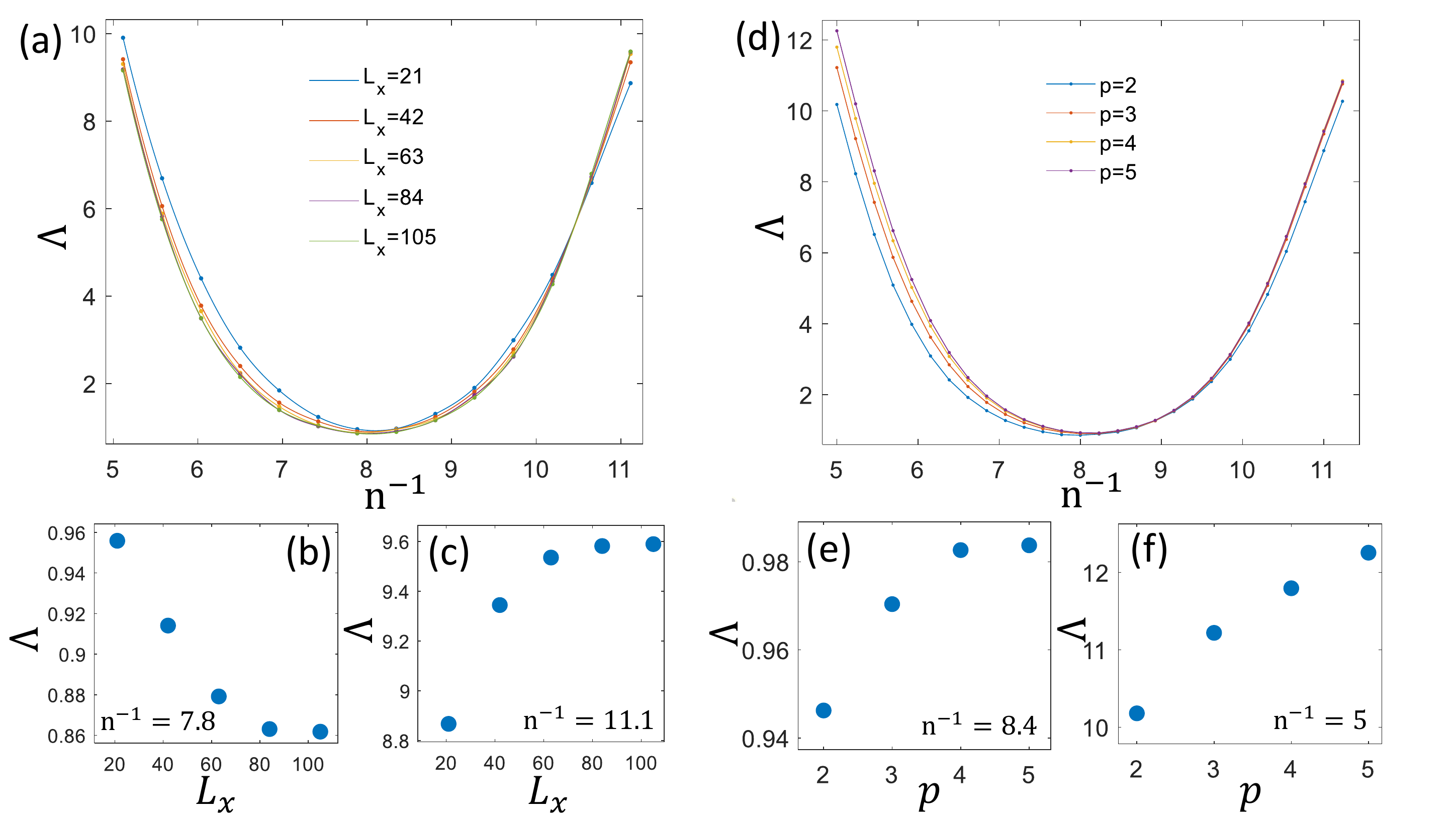}
\caption{\label{fig:Lx_p}(a) the dimensionless Lyapunov exponent $\Lambda = \tilde{\Lambda}L=L/\xi$, Eq.~(11) of the main text, as function of $n^{-1}$ for $L=28$, range cutoff $p=2$, and different values
of $L_{x}$ (the size in the $x$-direction of the $G^{-1}$ matrices).
(b), (c) $\Lambda$ as function of $L_{x}$, for two
specific values of $n^{-1}$ from panel (a).
(d) $\Lambda$ as function of $n^{-1}$ for $L=28$ and for different values of $p$
(the range cutoff). (e),(f) $\Lambda$ as function
of $p$, for two specific values of $n^{-1}$ from panel (d).}

\end{figure}

\red{
	\section{Parameter choice and universality of the critical exponent} \label{sec:universality}
}

\begin{figure}
	\includegraphics[scale=0.5]{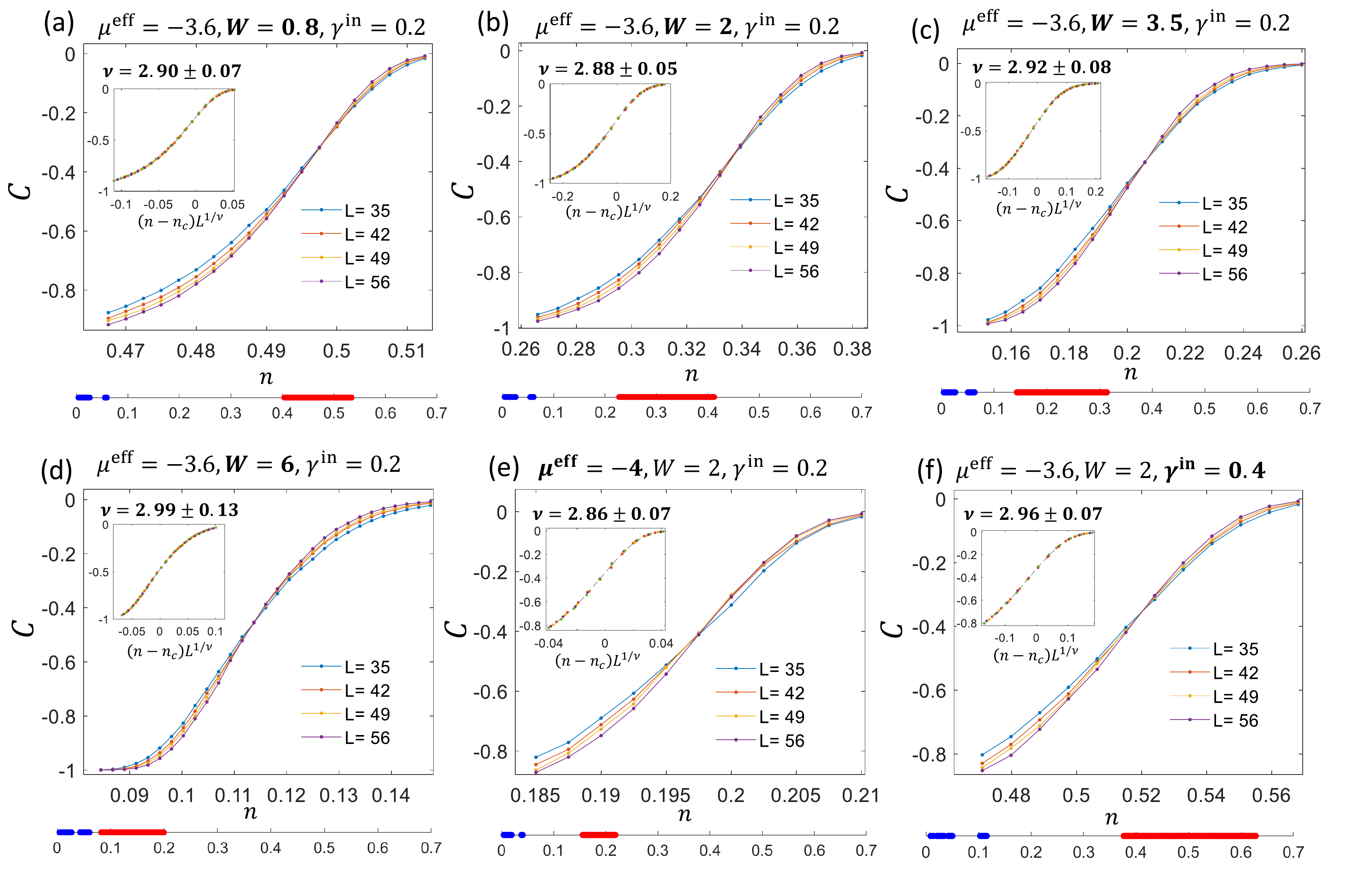}
	\centering{}\caption{\label{fig:local_chern_W}
		\red{
			The average local Chern marker out of equilibrium, and the critical exponent extracted using Method II
			for different parameter values. Panels (a-d): $\mu^{\mathrm{eff}}=-3.6$, $\gamma^{\mathrm{in}}=0.2$ and the disorder strength $W$ takes different values. Panel (e): $\mu^{\mathrm{eff}}=-4$, $W=2$, $\gamma^{\mathrm{in}}=0.2$. Panel (f):  $\mu^{\mathrm{eff}}=-3.6$, $W=2$, $\gamma^{\mathrm{in}}=0.4$.
			Insets: scaling data collapse.
			Bottom panel: the occupation bands. The critical exponent is seen to be universal, i.e., independent of the exact parameter values.}}
\end{figure}

\red{
	In equilibrium, our model contains two parameters: (i) The dimensionless magnetic
	flux through a unit cell, $\alpha=Ba^{2}/\phi_{0}$, where
	$\phi_{0}=h/e$ is the flux quantum and $a$ is the lattice
	spacing; (ii) The onsite disorder strength $W$. These values should
	be chosen employing the following considerations \citep{Puschmann2019}:
	(a) Unlike the continuum case, on the lattice each band of the Hofstadter model has an
	``intrinsic'' width $\delta$, the width of the band \textit{without}
	disorder. Therefore, $\alpha$ should be chosen such that $\delta/\Delta\ll1$,
	where $\Delta$ is the spacing between the Landau levels.
	However, a too small value of $\alpha$ is also not preferred since it
	would increase the magnetic length (measured in units of the lattice spacing $a$) $\ell_{B}=1/\sqrt{2\pi\alpha}$,
	making the effective system size $L/\ell_B$ smaller (one
	can compensate for this by working with larger system sizes $L$, but it
	would be expensive in terms of computation time). (b) The disorder
	strength should be large enough such that the disorder-induced broadening of the band would
	be much larger than $\delta$, but not as large as to mix between
	different Landau levels. If one picks the parameters following these
	considerations, one should obtain a universal value for the critical exponent, which
	is \textit{independent} of the exact parameter values \citep{Puschmann2019}.
	In our work, we have found $\alpha=1/7$ and $W=0.2$ to be appropriate
	in this respect.
}

\red{
	Out of equilibrium
	instead of energy bands we have ``occupation bands'', which are bands
	of eigenvalues $n$ (occupations) of the single-particle density matrix
	$G$. Without disorder in $H_S$ these bands are given by Eq.~(5) in the main text.
	Our nonequilibrium model contains four parameters: (i) $\alpha$
	of the reference Hamiltonian; (ii) $W$, the disorder strength in the system
	Hamiltonian $H_S$; (iii) $\mu^{\mathrm{eff}}$, the effective chemical potential; (iv) $\gamma^{\mathrm{in}}$, the refilling rate. We recall that we have
	proven in section \ref{subsec:Disorder-in-the} that if the system
	Hamiltonian $H_S$ is zero then different values of $\mu^{\mathrm{eff}}$
	and $\gamma^{\mathrm{in}}$ will not affect the critical exponent
	(as long as $\mu^{\mathrm{eff}}$ is not chosen near $E_{c}$, the
	critical energy of the band). While this proof does not hold for $W\neq0$, if the parameters are chosen by similar
	considerations to those presented above for the equilibrium case, their exact values will not change the
	result, as we will show in what follows.
	In this work we have taken $\alpha=1/7$, $\mu^{\mathrm{eff}}=-3.6$
	(that is, $E_{c}-\mu^{\mathrm{eff}}\approx0.4$, since $E_{c}\approx-3.2$), and $\gamma^{\mathrm{in}}=0.2$. For the disorder strength, we took $W=2$
	for Methods I and II. For Method III we chose $W=5.5$, since it results in a more symmetric behavior around $n_c$, which somewhat reduces the need for corrections to scaling.
}

\red{
	We will now present results which demonstrate that the critical exponent is indeed universal, in the sense of being insensitive to the exact parameter values. For concreteness we concentrate on Method II, though we have verified similar results hold for the other methods. We will separately change
	each one of the parameters while keeping the values of the rest the same. The results are plotted in Fig.~\ref{fig:local_chern_W}.
	In panels~\ref{fig:local_chern_W}(a-d) we compare
	the nonequilibrium scaling for different values of disorder $W$.
	We note that panel (b) corresponds to the same parameter values
	as in Fig.~2(b) in the main text, but also includes smaller system sizes.
	On the bottom panels we see the ``occupation bands'', the spectrum of the single-particle reduced density matrix $G$, which are in the range of 0 to 1 (since they represent occupation values). The band that we investigate is the one with highest occupations [since it would correspond to the lowest energy band in equilibrium, or even out of equilibrium when the disorder is in the system-bath coupling Hamiltonian, cf.~Eq.~(5) of the main text], which is marked in red. We note that here there are also 7 bands as in equilibrium, but bands 3-7 have occupations that are
	close to zero and therefore hard to resolve in the figure. In panel~\ref{fig:local_chern_W}(e) we take $\mu^{\mathrm{eff}}=-4$,
	which doubles the value of $E_{c}-\mu^{\mathrm{eff}}$ from $0.4$
	to $0.8$. In panel~\ref{fig:local_chern_W}(f) we take $\gamma^{\mathrm{in}}=0.4$
	instead of $0.2$. It is evident that while each different parameter choice leads to
	a change in the position and shape of the highest-occupancy band, all of the cases result
	in a similar value of $\nu\sim2.9$, agreeing with the result presented
	in the main text, and demonstrating their universality.
}



\section{Distribution of $G^{-1}$}
\label{sec:ginv}

\begin{figure}
	\begin{centering}
		\includegraphics[scale=0.68]{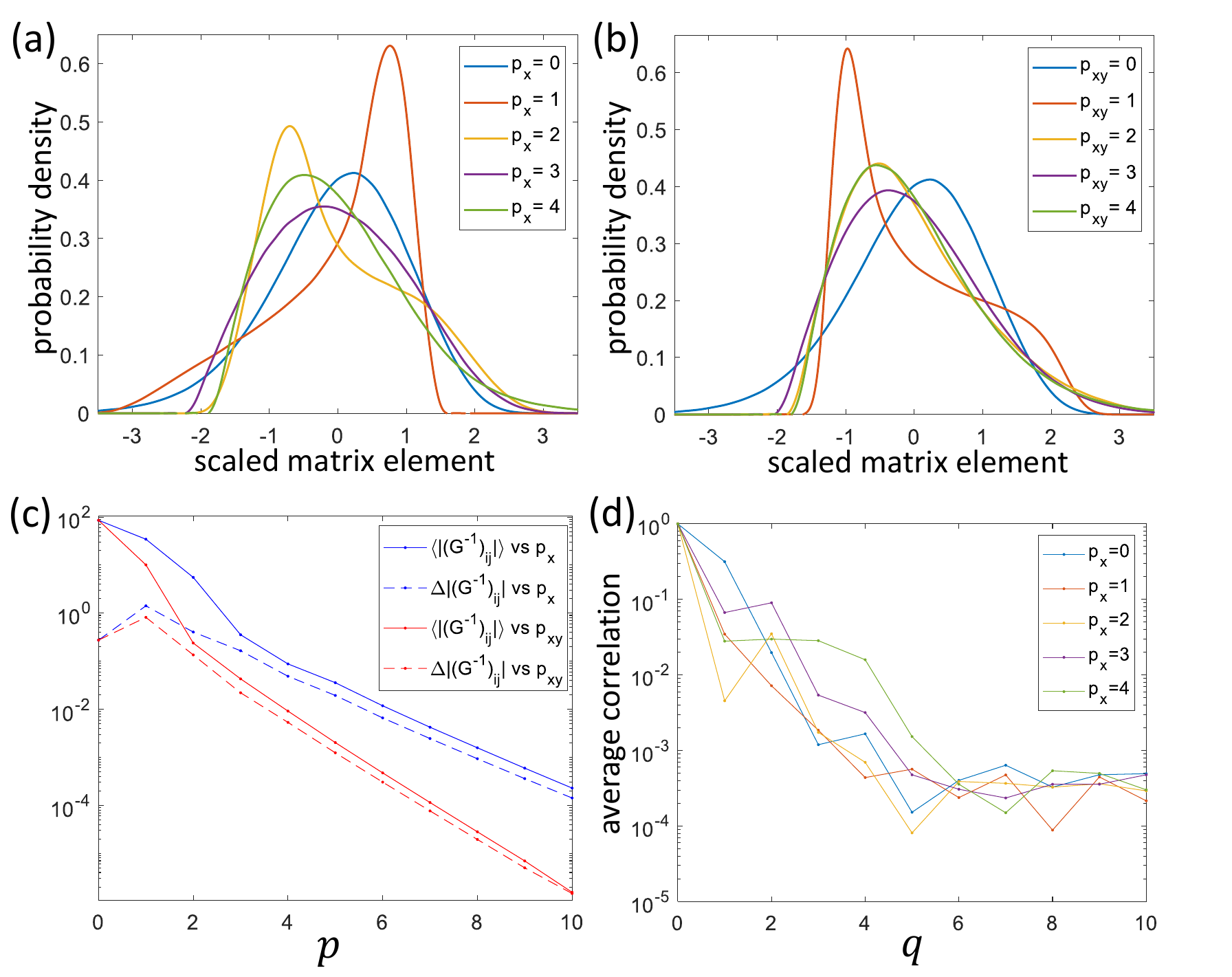}
		\par\end{centering}
	\caption{\label{fig:inv_G_distributions}
		(a) The distributions of absolute values of the matrix elements $(G^{-1})_{ij}$ connecting sites along the $x$-direction, where $i=(r_{x},r_{y})$ and $j=(r_{x}+p_{x},r_{y})$
		with $p_{x}=0,...,4$ ($p_{x}=0$ is the onsite term), shifted by their respective expectation values and normalized by their respective standard deviations.
		(b) The distributions of the absolute values of the matrix elements of $(G^{-1})_{ij}$ connecting sites along the $\pi/4$ diagonal in the $xy$-plane, where $i=(r_{x},r_{y})$ and $j=(r_{x}+p_{xy},r_{y}+p_{xy})$ with $p_{xy}=0,...,4$ ($p_{xy}=0$ is the onsite term), shifted by their respective expectation values and normalized by their respective standard deviations.
		(c) Semi-log plot of the expectation values and standard deviations of the absolute values of the elements $(G^{-1})_{ij}$ connecting
		sites along the $x$-direction (blue), and connecting sites along the $\pi/4$ diagonal in the $xy$-plane (red).
		(d)~Semi-log plot of the correlation $\mathcal{C}_{p_x}(q)$ [Eq.~(\ref{eqn:corr}, normalized by its value at $q=0$] of the absolute values of matrix elements $(G^{-1})_{ij}$ connecting sites along the $x$-direction with $p_{x}=0,...,4$. $q$ is the distance between the correlated terms (along either the $x$- or the $y$-direction).
		For the calculations, $M=2000$ realizations of $G^{-1}$ were generated, each with $L_{x}=105$, $L_{y}=28$ (with periodic boundary conditions), $\gamma^{\mathrm{in}}=0.2\gamma^{0}$, $\mu^{\mathrm{eff}}=-3.6$, and disorder strength $W=5.5$.}
\end{figure}

As discussed in the main text, out of equilibrium the steady state single-particle density matrix matrix $G$ plays the role of the Hamiltonian in characterizing both the topology of the system and its localization properties.
In that respect, concentrating on its inverse $G^{-1}$ offers some advantages.
This is particularly clear if $H_S=0$, that is, in the clean case or if disorder is included only in the reference Hamiltonian, since then $G^{-1}$ is simply related to $h^\mathrm{ref}$ via Eq.~(5) of the main text.
This implies that $G^{-1}$ and $h^{\mathrm{ref}}$ share the same eigenvectors, and moreover, that for nearest neighbor $h^{\mathrm{ref}}$, $G^{-1}$ has up to next-nearest neighbor terms ($p=2$), which allows its study via the transfer matrix without approximation with respect to the range.
However, as was mentioned in the main text, this is no longer the case when the disorder is included in the system Hamiltonian; now in order to obtain $G^{-1}$ one needs to solve numerically
the continuous Lyapunov equation, Eq.~(3) of the main text. 
Therefore, we will now study its statistical properties in this case.

Fig.~\ref{fig:inv_G_distributions}(a)--(b) presents:
(i) The distribution of the absolute values of onsite terms of the matrix $G^{-1}$, i.e., the diagonal terms $\left(G^{-1}\right)_{ii}$ where $i=(r_{x},r_{y})$;
(ii) The distribution of the absolute values of terms connecting sites along the $x$-direction, i.e., off-diagonal terms $\left(G^{-1}\right)_{ij}$ with $i=(r_{x},r_{y})$ and $j=(r_{x}+p_{x},r_{y})$ (similar distributions are obtained for sites separated along the $y$-direction);
(iii) The distribution of the absolute values of terms connecting sites along the $\pi/4$ diagonal in the $xy$-plane, i.e., the off-diagonal terms $\left(G^{-1}\right)_{ij}$ with $i=(r_{x},r_{y})$ and $j=(r_{x}+p_{xy},r_{y}+p_{xy})$.
The distributions (ii) and (iii) are shown, respectively, in panels (a) and (b); the onsite term (i) appears in both panels as the term $p_{x}=0$ and $p_{xy}=0$, respectively. The distributions are shifted by their respective expectation values and normalized by their respective standard deviations, which are presented in panel (c).
In panel (d), we can see the average correlation of the absolute values of the matrix elements of $G^{-1}$ which connect sites along the x-direction. The correlation is defined as:
\begin{equation} \label{eqn:corr}
	\mathcal{C}_{p_x}(q) = \left\langle \left|\left(G^{-1}\right)_{i_{1}i_{2}}\right|\left|\left(G^{-1}\right)_{j_{1}j_{2}}\right|\right\rangle -\left\langle \left|\left(G^{-1}\right)_{i_{1}i_{2}}\right|\right\rangle \left\langle \left|\left(G^{-1}\right)_{j_{1}j_{2}}\right|\right\rangle,
\end{equation}
where $i_{1}=(r_{x},r_{y})$, $i_{2}=(r_{x}+p_{x},r_{y})$, and where $j_1$ and $j_2$ are shifted with respect to $i_1$ and $i_2$ by $q$ sites along either the $x$-direction [$j_{1}=i_{1}+(q,0)$, $j_2 =i_{2}+(q,0)$] or the $y$-direction [$j_{1}=i_{1}+(0,q), j_{2} =i_{2}+(0,q)$] --- both options gave similar results, and we averaged over them to reduce statistical noise. The correlation is normalized by its value at $q=0$ (the variance), which can be inferred from the standard deviations plotted in panel (c).

An important observation is that the distributions do not feature any long tails.
Moreover, Fig.~\ref{fig:inv_G_distributions}(c) shows that both the expectation values and standard deviations of the various terms decay exponentially with range ($p_{x}$ or $p_{xy}$). And Fig.~\ref{fig:inv_G_distributions}(d) demonstrates that the same is true for the correlations between different elements (the saturation at $q>6$ is due to the values becoming smaller than the statistical error).
This justifies cutting off the range, as done in the transfer matrix method III out of equilibrium. Moreover, as notes in the main text, the new localization universality class we find cannot be attributed to $G^{-1}$ having terms with long range or fat-tailed distributions, and therefore seems to be a genuine nonequilibrium effect.

Finally, a curious fact is that one can derive an analytic result for the expectation value of the onsite terms. Starting from the continuous Lyapunov
equation [Eq.~(3) of the main text], 
we can multiply by $G^{-1}$ from the right and then take the trace, leading to
\begin{equation}
	\gamma^{\mathrm{in}}\mathrm{Tr}\left(G^{-1}\right)=\mathrm{Tr}\left(\gamma^{\mathrm{in}}+\gamma^{\mathrm{out}}\right),
\end{equation}
independently of $h_{S}^{*}$. Taking $h^{\mathrm{ref}}$ to be the Hofstadter Hamiltonian [Eq.~(4) of the main text], we
can see that $\mathrm{Tr}(h^{\mathrm{ref}})=0$ and $\mathrm{Tr}\left(\left(h^{\mathrm{ref}}\right)^{2}\right)=4L_{x}L_{y}$.
Substituting $\gamma^{\mathrm{out}}=\gamma^{0}\left(h^{\mathrm{ref}}-\mu^{\mathrm{eff}}\right)^{2}$, this results in
\begin{equation} \label{eqn:ginv_ii_av}
	\langle(G^{-1})_{ii}\rangle=1+\frac{1}{\gamma^{\mathrm{in}}}(4+\mu^{*2}),
\end{equation}
since $\langle(G^{-1})_{ii}\rangle$ is independent of $i$ (for periodic boundary conditions), and hence equals $\mathrm{Tr}(G^{-1})/L_{x}L_{y}$. 
Eq.~(\ref{eqn:ginv_ii_av}) shows that the onsite average depends only on $\gamma^{\mathrm{in}}$ and $\mu^{\mathrm{eff}}$, but is independent of the disorder strength $W$; this would no longer be true for the corresponding standard deviation.


\end{widetext}

\end{document}